\documentclass[a4paper,11pt]{article}
\usepackage{jinstpub} % for details on the use of the package, please see the JINST-author-manual
\usepackage{lineno}
\usepackage{subcaption}
\usepackage{xcolor}  % package for colored text
\usepackage{placeins}

\title{\boldmath Gain characterization of LGAD sensors with beta particles and 28-MeV protons}

\author[a]{M. H. Mohamed Farook,}
\author[c]{G. Giacomini,}
\author[b]{G. D'Amen,}
\author[c]{G. Pinaroli,}
\author[c]{E. Rossi,}
\author[a]{S.~Seidel,}
\author[b,1]{and A. Tricoli\note{Corresponding author.}}
\affiliation[a]{Department of Physics and Astronomy, University of New Mexico, Albuquerque, New Mexico, USA}
\affiliation[b]{Physics Department, Brookhaven National Laboratory, Upton, New York, USA}
\affiliation[c]{Instrumentation Department, Brookhaven National Laboratory, Upton, New York, USA}

\emailAdd{atricoli@bnl.gov}

\abstract{Low Gain Avalanche Diodes, also known as LGADs, are widely considered for fast-timing applications in high energy physics, nuclear physics, space science, medical imaging, and precision measurements of rare processes. Such devices are silicon-based and feature an intrinsic gain due to a $p{^+}$-doped layer that allows the production of a controlled avalanche of carriers, with multiplication on the order of 10-100.
This technology can provide time resolution on the order of 20-30 ps, and variants of this technology can provide precision tracking too. The characterization of LGAD performance has so far primarily been focused on the interaction of minimum ionizing particles for high energy and nuclear physics applications. This article expands the study of LGAD performance to highly-ionizing particles, such as 28-MeV protons, which are relevant for several future scientific applications, e.g.\ in biology and medical physics, among others. These studies were performed with a beam of 28-MeV protons from a tandem Van de Graaff accelerator at Brookhaven National Laboratory and beta particles from a $^{90}{\rm Sr}$ source; these were used to characterize the response and the gain of an LGAD as a function of bias voltage and collected charge. The experimental results are also compared to TCAD simulations.}

\keywords{Solid state detectors, Particle tracking detectors, Timing detectors, protons, electrons}

\arxivnumber{2503.07941} % Only if you have one

\begin{document}
\maketitle
\flushbottom

%%%%%%%%%%%%%%%%%%%%%%%%%%%%%%%%%%%%%%%%%%%%%%%%%
\section{Introduction}
\label{sec:intro}
%%%%%%%%%%%%%%%%%%%%%%%%%%%%%%%%%%%%%%%%%%%%%%%%%

%For internal references use label-refs: see section~\ref{sec:intro}. Bibliographic citations can be done with "cite": refs.~\cite{a,b,c}. When possible, align equations on the equal sign. The package

Fast-timing detectors have attracted widespread interest for their applications to the development of the next generation of colliders in high energy physics experiments and for other applications including nuclear physics, space science, medical imaging, and precision measurements of rare processes. In this context, silicon-based particle sensors with fast-timing capabilities on the order of \mbox{30 ps} are a rapidly developing research area, for application in timing systems for the upgraded ATLAS and CMS experiments~\cite{CMS:2667167,Collaboration:2623663} at the CERN Large Hadron Collider (LHC) for the High Luminosity phase (HL-LHC)~\cite{CERN-ACC-2015-0140,CERN-ATS-2012-236}. One of the selected technologies for such detectors is a silicon pad with internal amplification, i.e.\ the Low Gain Avalanche Diode (LGAD)~\cite{hartmut,GIACOMINI201952,micronlgad}.
The principle of operation of an LGAD is based on the creation of a thin layer of high electric field at the $n^{++}$ and $p^{+}$ junction where the signal is amplified by avalanche multiplication to achieve a gain of about 10-100. The LGADs for the HL-LHC are implemented as a 50 $\mu$m thick $p$-type bulk silicon sensor with a highly doped $p^{+}$ layer, called the gain layer, under the $n^{++}$ electrode. Such sensors have also been fabricated recently on a thinner $p$-type active substrate, e.g.\ 20, 30, 33, 35 or 45 $\mu$m, to achieve faster time performance~\cite{Lange:2017pxs, Zhao:2018qkg,Li:2021iid,YANG2020164379,DUTTA2025170224}. Variants of the LGAD technology, e.g. the AC-coupled LGADs (AC-LGADs)~\cite{ACLGADprocess,RSD_NIM}, have been proposed for four-dimensional detectors in nuclear physics experiments, such as the ePIC experiment at the Electron-Ion Collider (EIC)~\cite{AbdulKhalek:2021gbh}, since they provide time resolution similar to that of LGADs and segmentation as small as a few tens of microns with a fill factor of 100$\%$. %While this article focuses on the LGAD performance, the general features of the results are expected to hold also in the case of AC-LGADs.

So far, the characterization of LGAD performance has focused primarily on the interaction of minimum ionizing particles (MIPs) for high energy physics and nuclear physics applications. This article expands the study to highly-ionizing particles (non-MIPs), which are relevant for future scientific applications in biology, medical physics, and precision measurements of rare processes, where low energy protons, pions, kaons or heavy ions are used. Among such applications, it is worth highlighting those in micro-dosimetry for radiotherapy, where low energy protons ($\leq 200$~MeV beam energy) are used as well as heavy-ion beams such as carbon~\cite{10.3389/fphy.2020.578444}. Also relevant is the PIONEER experiment at PSI~\cite{instruments5040040}, where low energy electrons and muons are detected to measure the ratios of the decay rates of charged pions to prompt electrons and muons. 

While MIPs produce about 60-80 electron-hole pairs per micron in thin silicon sensors, non-MIPs can produce a far greater amount of charge in silicon, and this may affect the signal properties of LGADs due to different charge multiplication and collection processes.
This article reports the study of LGAD signal properties performed with a beam of 28-MeV protons (which are non-MIP) from a tandem Van de Graaff accelerator at Brookhaven National Laboratory~\cite{Tandem}. The results are compared to those obtained with a beam of beta particles (MIPs) from a $^{90}{\rm Sr}$ source and to TCAD simulations to facilitate their interpretation. The gain and other properties of an LGAD are measured as a function of bias voltage and compared to those of diodes, i.e.\ sensors designed and fabricated with the same process and geometry as the LGAD but without the $p^+$-doped gain layer.

%%%%%%%%%%%%%%%%%%%%%%%%%%%%%%%%%%%%%%%%%%%%%%%%%
\section{Experimental apparatus}
\label{sec:DUT}
%%%%%%%%%%%%%%%%%%%%%%%%%%%%%%%%%%%%%%%%%%%%%%%%%

\begin{figure}[htbp]
    \centering
    \includegraphics[width=.5\linewidth]{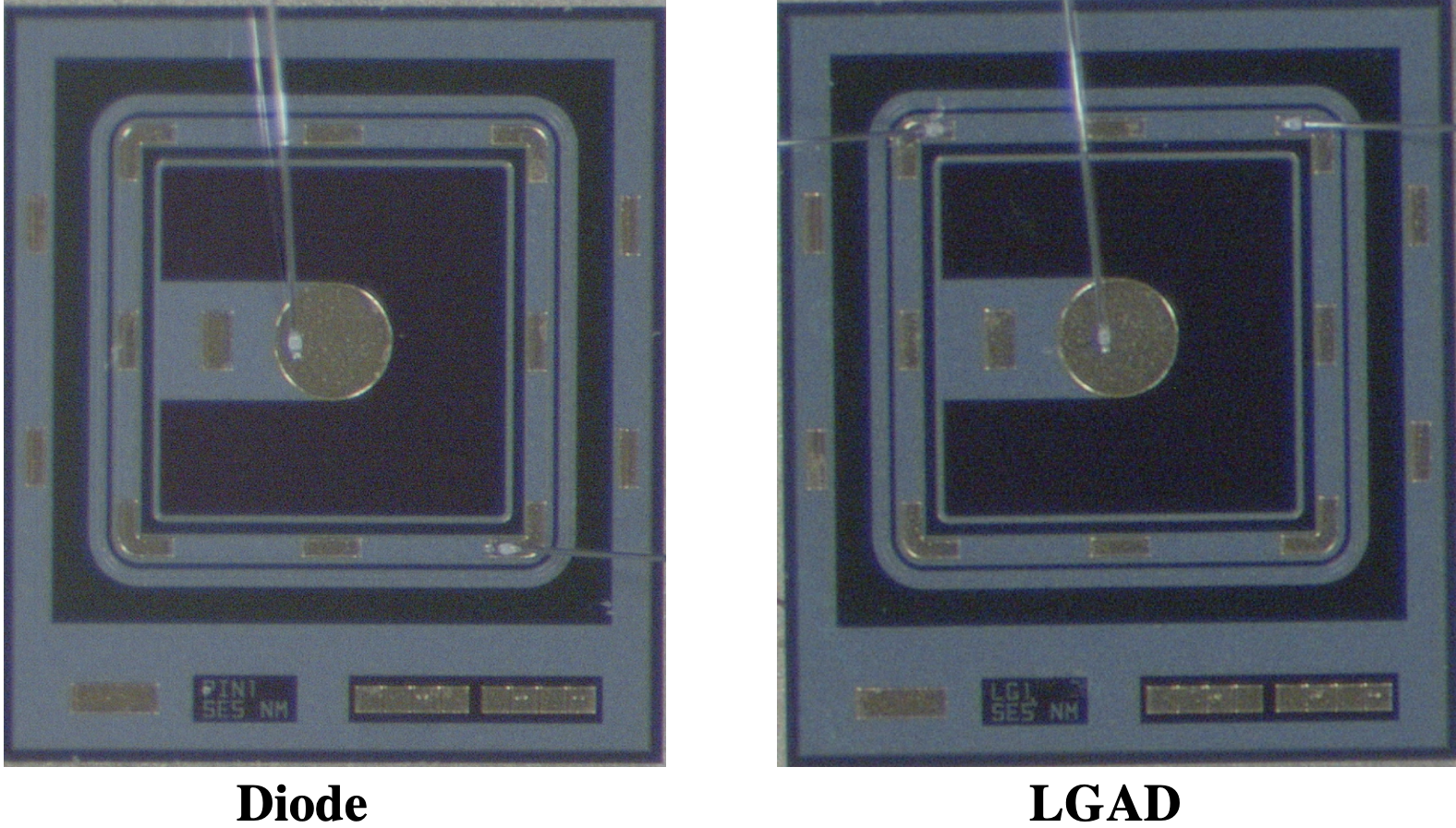}
     \includegraphics[width=.5\linewidth]{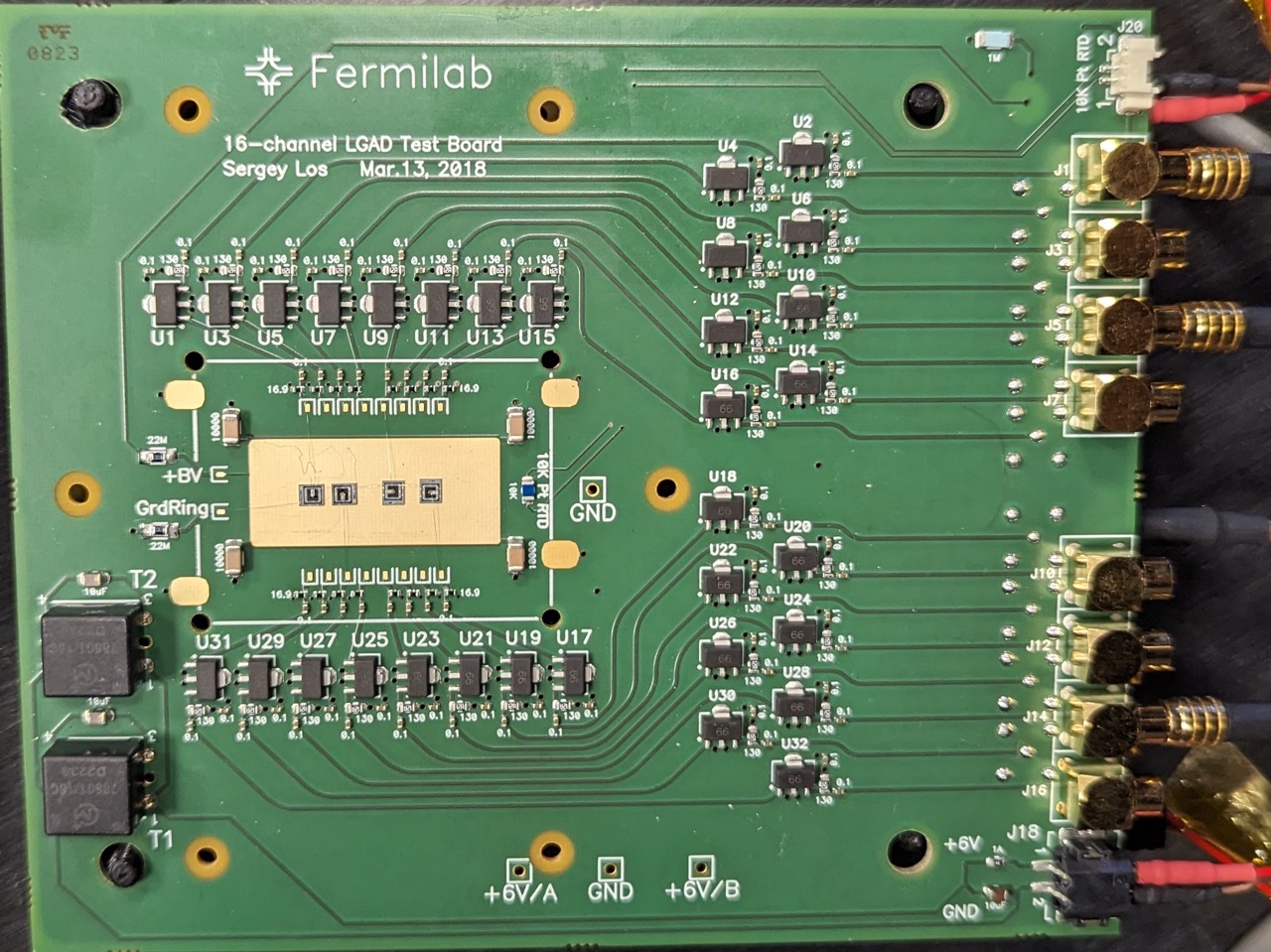}
    \caption{A diode (top-left) and the LGAD (top-right) sensors fabricated by HPK. The photos also shows the wire-bonds connected to the pads and the guard-rings. The FNAL readout board (bottom) is shown with sensors mounted and wire-bonded. (One additional sensor is mounted on the board, but it was found not to be well functioning and was not used for this study.)}
    \label{fig:sensors}
\end{figure}

Three silicon sensors were used for this study. They were fabricated by Hamamatsu Photonics (HPK): one LGAD of HPK type 1.2~\cite{HELLER2021165828}, and two diodes (Diode-1 and Diode-2). Two diodes were used for systematic checks.
All the sensors are single-pad of size 1.3 $\times$ 1.3 mm$^2$ and active thickness 35 $\mu$m. Figure~\ref{fig:sensors} (top) shows one diode and the LGAD. The LGAD is fully depleted at about 35-36 V. At operating bias voltage $\ge 100$ V the detector capacitance is about 5.5 pF. The LGAD current starts to go into the breakdown region at a bias voltage of about 250~V, and at bias voltages in the range $100-250$ V the leakage current is approximately $4.7-7.1$ nA. 

The three sensors were mounted on the same readout board, which was designed by FNAL, see Figure~\ref{fig:sensors} (bottom). It is a 16-channel board with two-stage amplification based on the Mini-Circuits GALI-66+ integrated circuit~\cite{MiniCircuitsGALI66}. The amplifiers use 50 $\Omega$ input impedance, for a total transimpedance of approximately 5 k$\Omega$ and a bandwidth of 1 GHz with gain of about 10 and up to 3 GHz with lower gain~\cite{HELLER2021165828}.

The diode and LGAD pulses were recorded using a four-channel Lecroy Waverunner 9404M-MS oscilloscope with a bandwidth of 1 GHz and a sampling rate of 10 GS/s per channel.

%%%%%%%%%%%%%%%%%%%%%%%%%%%%%%%%%%%%%%%%%%%%%%%%%
\section{Particle beams and data selection}
\label{sec:beams}

To characterize the performance of the sensors with non-MIPs, a beam of 28-MeV protons was used. Beta particles from a $^{90}{\rm Sr}$ radioactive source were used for MIP interactions. The stopping power of a 28-MeV proton is 15.5 MeV cm$^2$/ g. For reference, that for a 2.25-GeV proton (a MIP) is  1.7 MeV cm$^2$/ g, which is numerically very close to the stopping power of a 2.28-MeV beta-particle from the decay chain of a $^{90}{\rm Sr}$ radioactive source, which  is 1.6  MeV cm$^2$/ g~\cite{NISt}. 
%Therefore, the energy deposited in silicon by a 28-MeV proton is one order of magnitude greater than that of a MIP. 

%%%%%%%%%%%%%%%%%%%%%%%%%%%%%%%%%%%%%%%%%%%%%%%%%

%%%-----------------------
\subsection{Beta $^{90}{\rm Sr}$ source}
\label{sec:beta}

The ${}^{90}$Sr radioactive source was used with a 1 mm thick  aluminum layer placed in front to absorb low energy beta particles and select the 2.28-MeV ones from the decay chain. The radioactive source, placed about a centimeter from the device under test, illuminated the junction side of the devices. %see Figure~\ref{fig:Sr90_source_over_the_board}. 
The entire setup was enclosed in a metallic light-proof box for electromagnetic shielding. The measurements were taken at room temperature. The lowest trigger level used for the oscilloscope  to record events was 8 mV.
Events recorded by the oscilloscope were rejected if they were compatible with pick-up noise in the readout electronics.
%
%\begin{figure}[htbp]
%   \centering
%    \includegraphics[width=.4\linewidth]{figures/4_Sr90_on_board.png}
%    \caption{The $^{90}$Sr source of beta particles placed over the FNAL readout board on which the sensors are mounted.}
%    \label{fig:Sr90_source_over_the_board}
%\end{figure}
%
\\
The ${}^{90}$Sr radioactive source was not used for the diodes because the signals are very small and the triggering threshold, after tuning to reject the noise, cut the peak of the amplitude spectrum.

%%%----------------------------
\subsection{28-MeV proton beam}
\label{sec:protons}

The tandem Van de Graaff Facility at BNL consists of two 15-MV electrostatic accelerators capable of delivering continuous or high-intensity pulsed ion beams to experimental chambers in a wide range of ion species at various energies~\cite{Tandem}. For this experiment a beam of 28-MeV protons was used with a flux in the range 1-5 $\times 10^4$ protons/(s cm$^2$). Data were recorded for between 5 minutes and 40 minutes per working point. At least 10,000 events were collected for diodes and for the LGAD at bias voltages below 200 V, and at least 80,000 events were collected for the LGAD at higher voltages.
The FNAL readout board with the sensors was placed inside a vacuum chamber that is located at the end of the beam line. The board was connected through a thermal glue to a cold chuck that kept the temperatures stable at about $25^\circ$C using a chiller that ran with a fluid consisting of a mix of water and glycol. High voltage to the devices under test was provided by a Keithley 2410 as a power supply unit which was placed outside the vacuum chamber and controlled remotely by the operators in a control-room through a custom-made interface based on Python drivers. 

\begin{figure}[htbp]
    \centering
    \includegraphics[width=.6\linewidth]{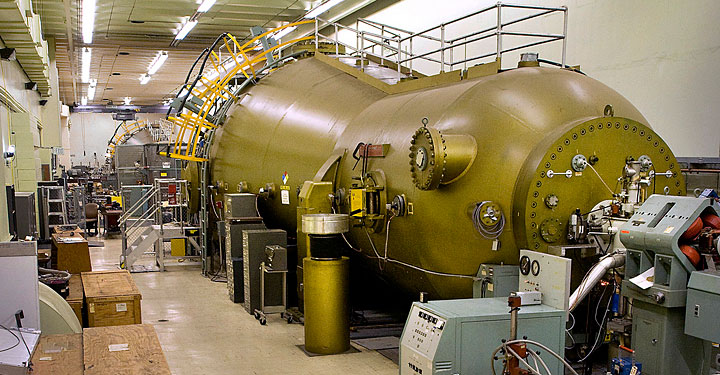}
    \includegraphics[width=.35\linewidth]{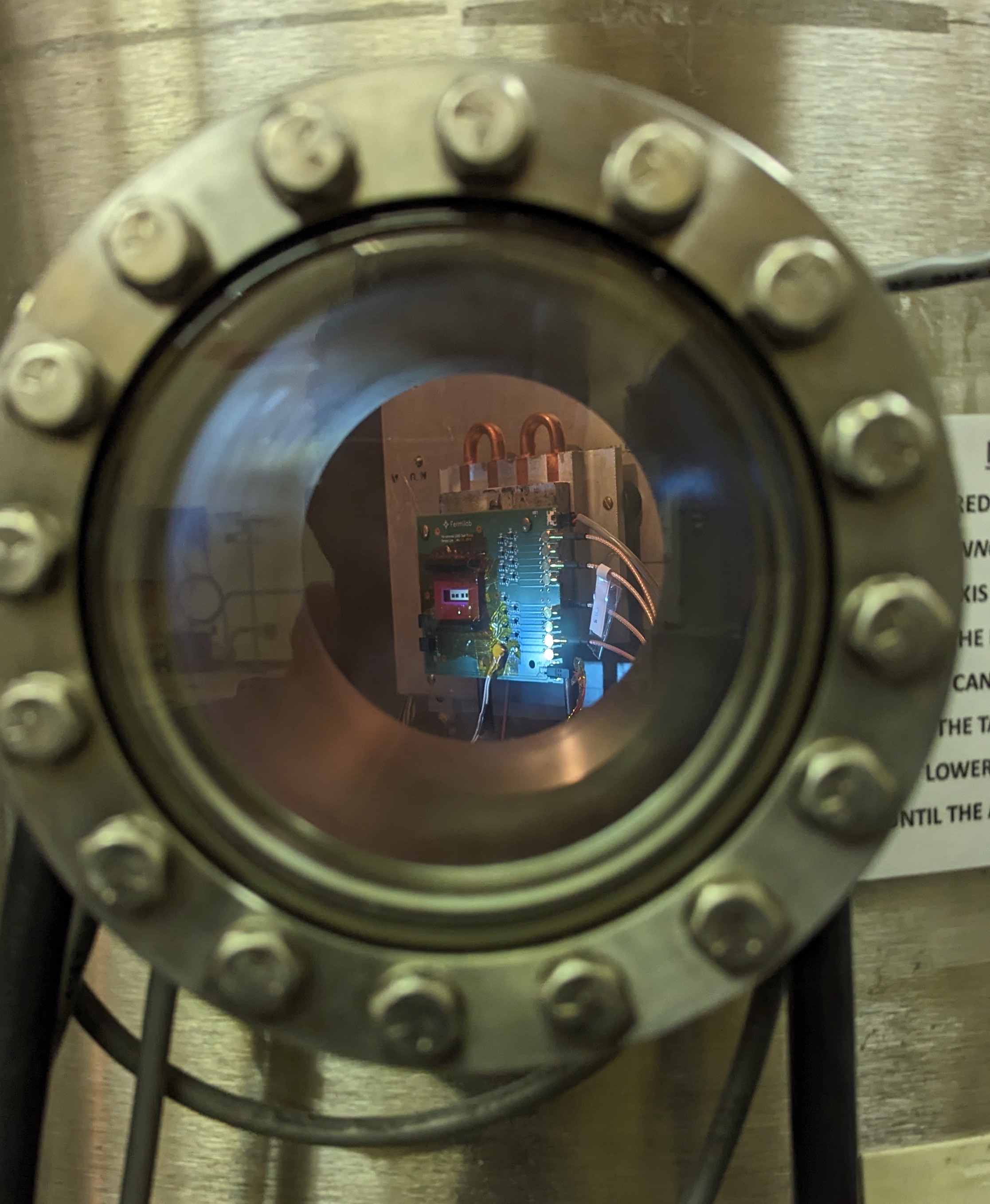}
    \caption{The tandem Van de Graaff (left) and the devices under test inside the vacuum chamber (right).}
    \label{fig:Tandem}
\end{figure}

The noise level in each electronic channel (one channel corresponding to one sensor) was initially measured without beam. The average baseline noise level was confirmed with beam on an event-by-event basis as the mean of the voltage in the readout time-window prior to the signal pulse. The baseline noise was subtracted on an event-by-event basis from the measured pulse. The standard deviation of the noise was found to be in the range 1.12 - 1.33 mV independently of the bias voltage. The trigger levels in the oscilloscope were adjusted according to the noise level in each channel and the oscilloscope voltage scale and were set in the range  5.1$\sigma$ to 19.1$\sigma$ above the noise.
Pulses with amplitudes above 1.2 V or above the oscilloscope voltage range were rejected as they saturate the readout electronics. This selection affected only the LGAD and rejected a small fraction of events, with no significant impact on the selected data set.  
Events recorded by the oscilloscope were rejected if they were compatible with pick-up noise in the readout electronics.

%-Base line is always considered to be 0, this is done by finding the noise by estimating the mean voltage before the signal pulse and subtracting the waveform by the mean voltage.\\
%-Initially noise was recorded without injecting beam on each channel with different bias voltages. Sigma of the noise for each channel was as follows:\\
%Channel 1 (LGAD1) = 1.33 mV  \\
%Channel 2 (Diode1)= 1.13 \\
%Channel 3 (LGAD2) = 1.15 mV  \\
%Channel 4 (Diode2)= 1.12,\\
%this was independent of the bias voltage of the DUT. \\
%-Trigger levels were adjusted according to each bias voltage. This was because the oscilloscope scale was adjusted to capture the full signal. As the voltage scale was increased the trigger level was also adjusted to trigger the events (This was due to some limitation of the oscilloscope). \\
%-Only the waveforms within the oscilloscope set voltage range were selected (saturated waveforms were excluded) \\
%All waveforms with maximum amplitude more than 1.2 V were excluded (limitaition of the FNAL board) \\
%%%%
%-Final cleanup of the waveform was performed by plotting the distribution of FWHM vs Ampl of all the waveforms and we were able to identify demarcated regions of signal and noise/gain less pulses. 

%%%%%%%%%%%%%%%%%%%%%%%%%%%%%
\subsection{Signal pulses in data}
\label{sec:signal_wf}

\begin{figure}[htbp]
% \centering
      \setkeys{Gin}{width=\linewidth}
     \begin{subfigure}[t]{0.32\textwidth}
         \includegraphics{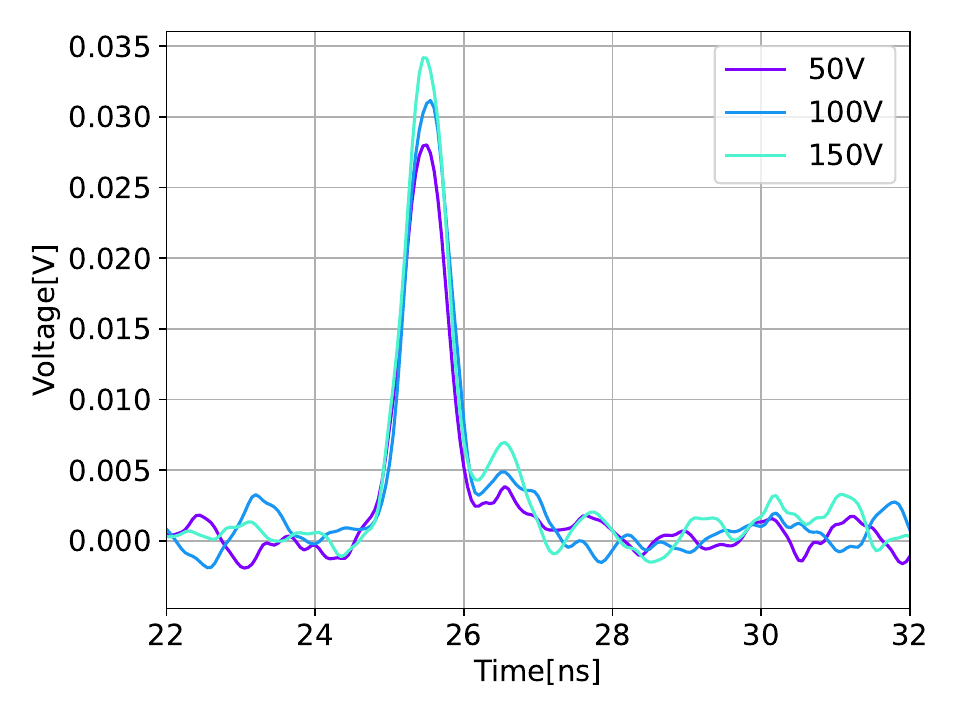}%{Picture1}
         \caption{Diode with 28-MeV proton}
         \label{fig:waveform_proton_diode}
     \end{subfigure}
     \hfill
     \begin{subfigure}[t]{0.32\textwidth}
         \centering
         \includegraphics{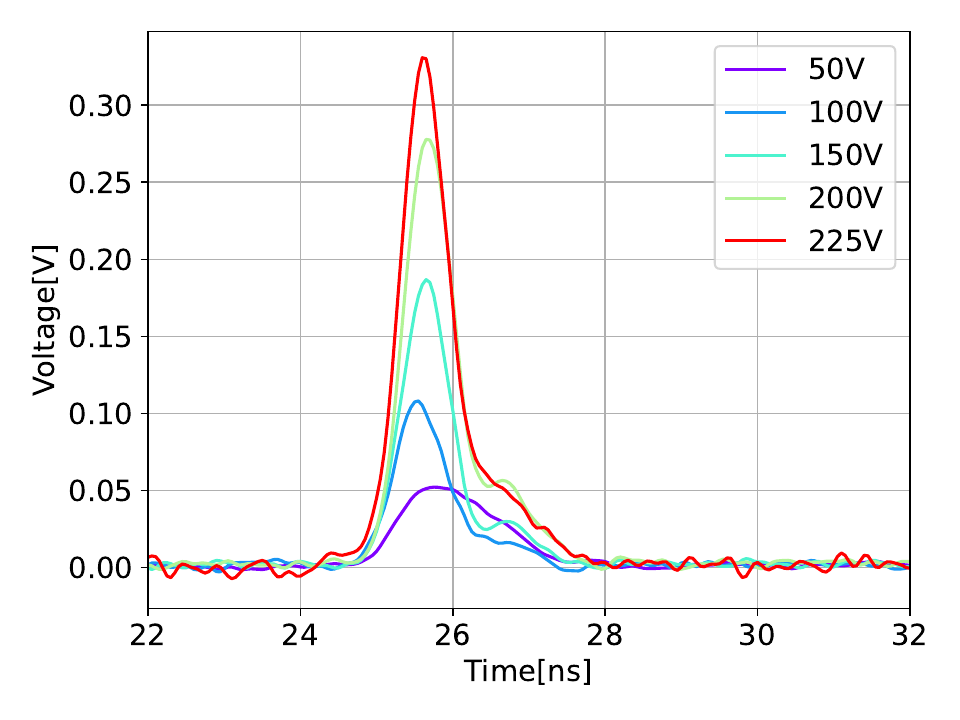}%{Picture2}
         \caption{LGAD with 28-MeV proton}
         \label{fig:waveform_proton_lgad}
     \end{subfigure}
     \hfill
     \begin{subfigure}[t]{0.32\textwidth}
         \centering
         \includegraphics{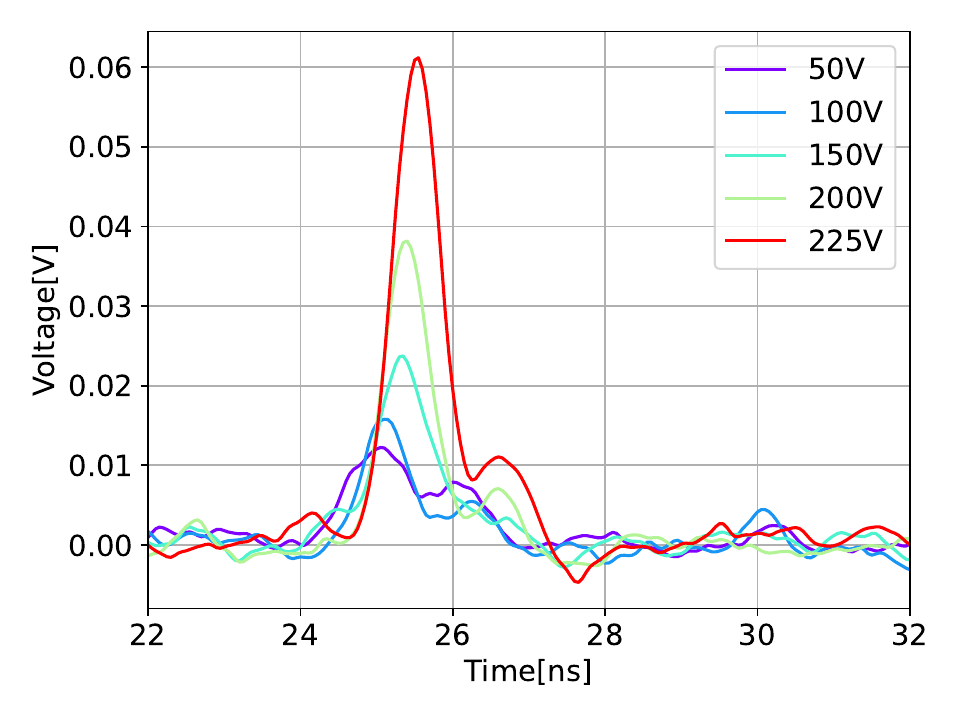}%{Picture3}
         \caption{LGAD with beta-particle}
         \label{fig:waveform_proton_lgad}
     \end{subfigure}
     \caption{Examples of typical pulses generated by Diode-1 in a beam of 28-MeV protons (a), and by the LGAD in a beam of 28-MeV protons (b) and beta-particles (c) at different values of bias voltage.}
    \label{fig:waveforms_data}
\end{figure}

Examples of typical pulses from the LGAD, recorded using both 28-MeV protons and beta-particles, and those from Diode-1 for 28-MeV protons, are shown in Figure~\ref{fig:waveforms_data}. As explained in the previous section, for diodes, pulses from beta-particles are not shown as the signal is small with respect to the electronic noise, and small-amplitude signals are rejected by the trigger threshold set on the oscilloscope, creating a bias on the selected pulses. To measure the signal amplitudes in diodes with MIPs, a different method is followed, see Section~\ref{sec:gain_ampl_beta}.  
For the LGAD, pulses are shown in Figure~\ref{fig:waveforms_data} for the bias voltage set in the range 50-225 V. For the diode, only pulses for bias voltages up to 150 V are shown. In the diode, the pulses show a small dependence on the bias voltage and do not change significantly above 150 V. In the LGAD case, the signal amplitudes are one order of magnitude greater than those in the diodes, and the increase in amplitudes as the bias voltage increases is noticeable.  In addition, with a 28-MeV proton beam, the LGAD amplitude is greater than that from beta-particles by up to a factor of 8 (at 150 V bias voltage).

%%%---------------------------

%%%%%%%%%%%%%%%%%%%%%%%%%%%%%%%%%%%%%%%%%%%%%%%%%
\section{Signal properties} 
\label{sec:signal}
%%%%%%%%%%%%%%%%%%%%%%%%%%%%%%%%%%%%%%%%%%%%%%%%%

To characterize the response of the LGAD to non-MIP interactions from 28-MeV protons, several key parameters are studied, such as the pulse amplitude, pulse area (the integral of the pulse as a function of time), the full-width at half-maximum (FWHM), and the rise time. The latter parameter is defined as the difference between the times when the pulse surpasses 40$\%$ and $90\%$ of its peak value. Such threshold values were optimized to minimize the effect of noise in the rise time measurement.  %Gains for the LGAD is also calculated. 
These quantities are compared to those measured with beta-particles in LGADs as well as those measured with 28-MeV protons in diodes.

%%%---------------------------
\subsection{Signal properties in diodes}
\label{sec:signal_diodes}

\begin{figure}[t]
% \centering
      \setkeys{Gin}{width=\linewidth}
     \begin{subfigure}[t]{0.45\textwidth}
         \includegraphics{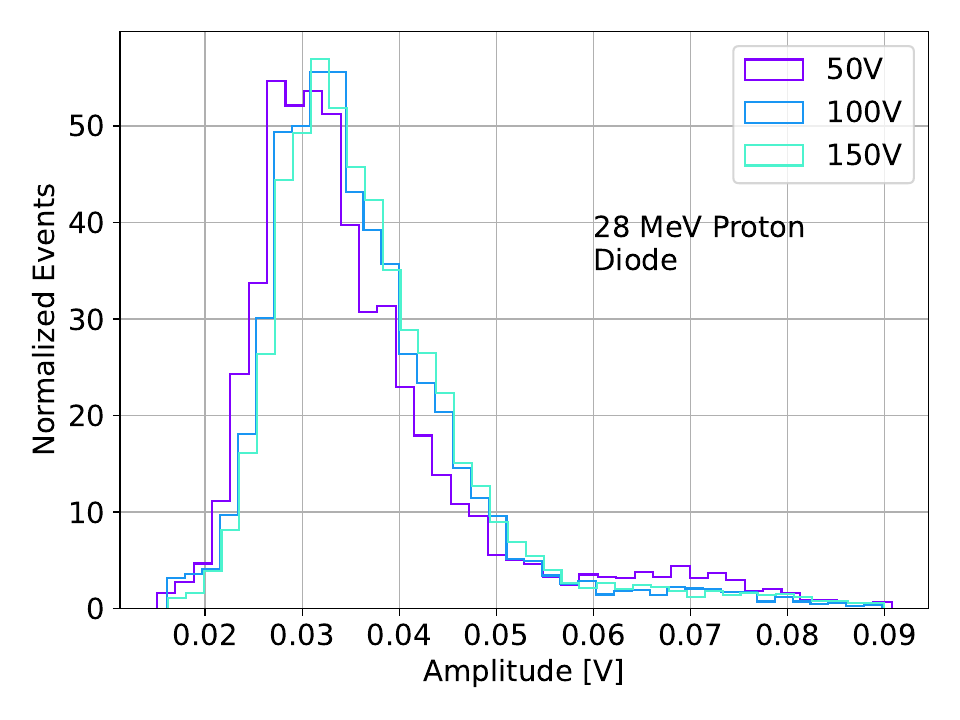}%{Picture1}
         \caption{}
         \label{fig:Diode_amplitude}
     \end{subfigure}
     \hfill
          \begin{subfigure}[t]{0.47\textwidth}
         \centering
         \includegraphics{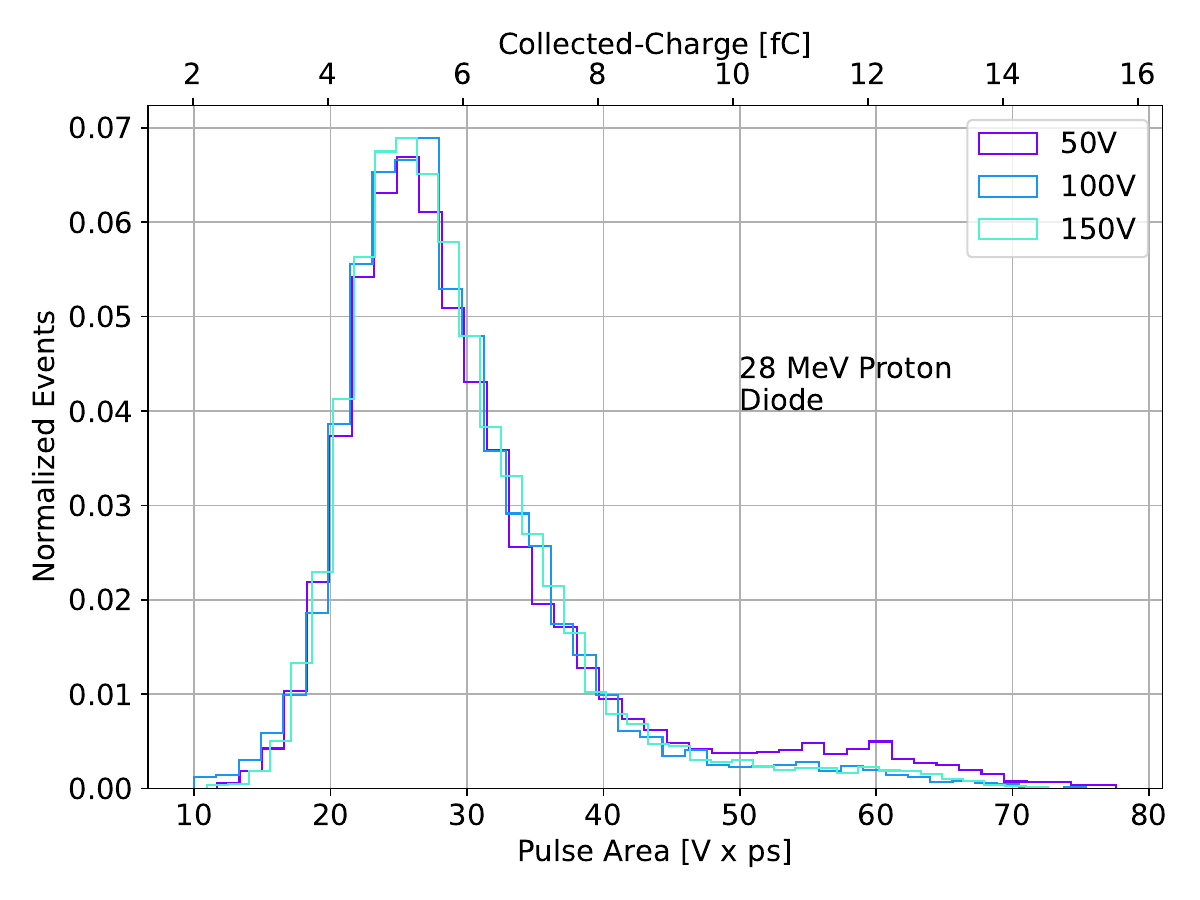}%{Picture3}
         \caption{}
         \label{fig:Diode_Area}
         \end{subfigure}
     \hfill
     \begin{subfigure}[t]{0.45\textwidth}
         \centering
         \includegraphics{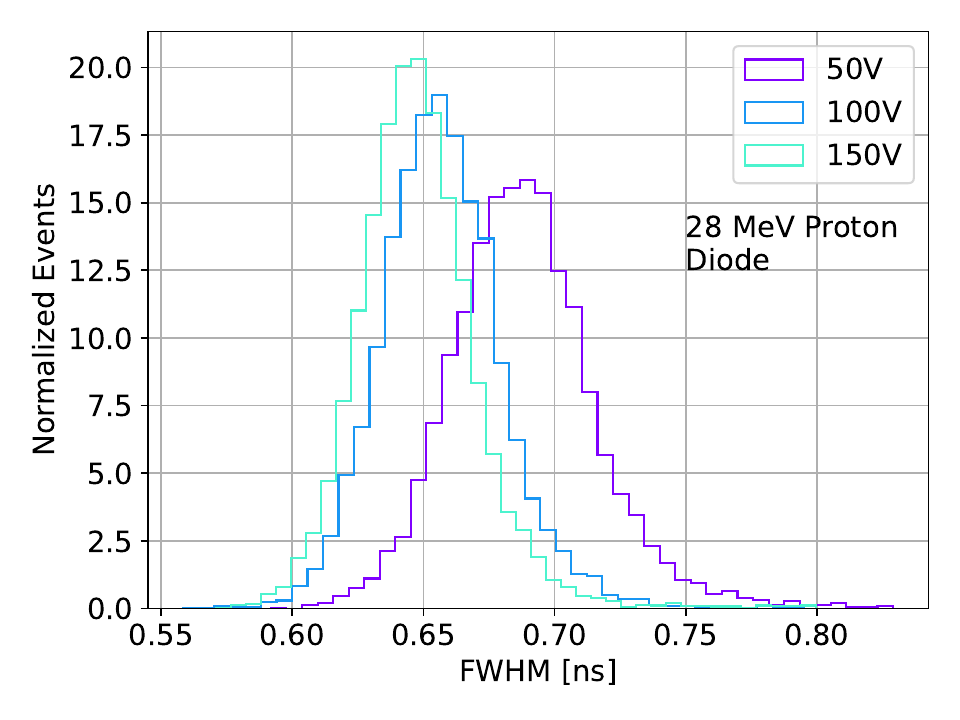}%{Picture2}
         \caption{}
         \label{fig:Diode_FHWM}
     \end{subfigure}
     \hfill
     \begin{subfigure}[t]{0.45\textwidth}
         \centering
         \includegraphics{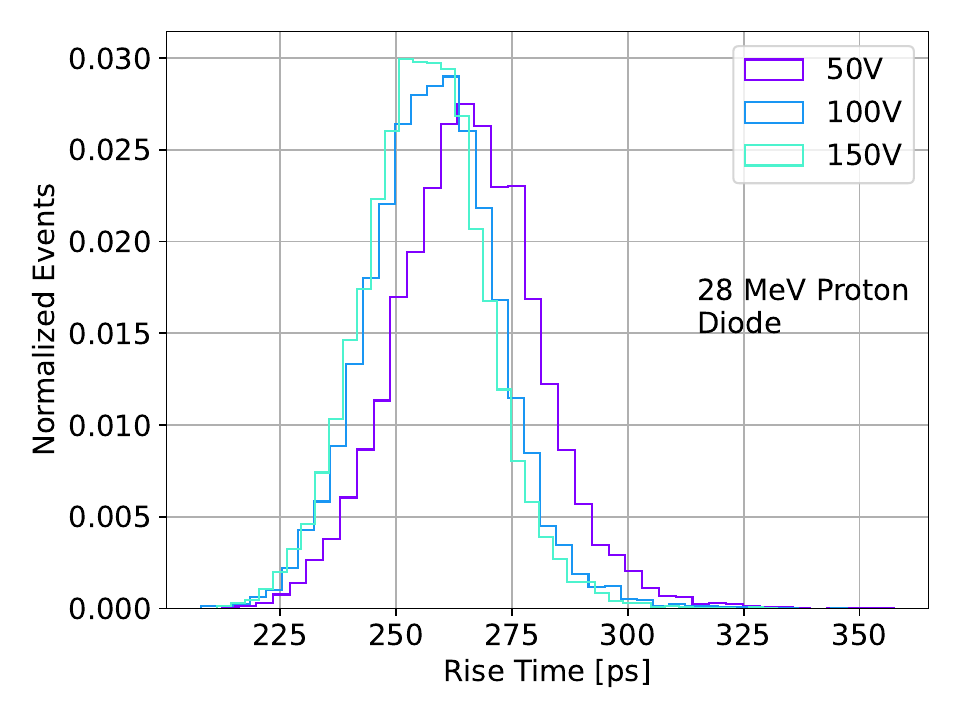}%{Picture3}
         \caption{}
         \label{fig:Diode_RiseTime}
     \end{subfigure}
     \caption{Area-normalized distributions of amplitude (a), area (b), FWHM (c) and rise time (d) for Diode-1 in a beam of 28-MeV protons as functions of bias voltage.}
    \label{fig:signal-prop_diode_28MeVproton}
\end{figure}

\begin{figure}[t]
% \centering
      \setkeys{Gin}{width=\linewidth}
     \begin{subfigure}[t]{0.45\textwidth}
         \includegraphics{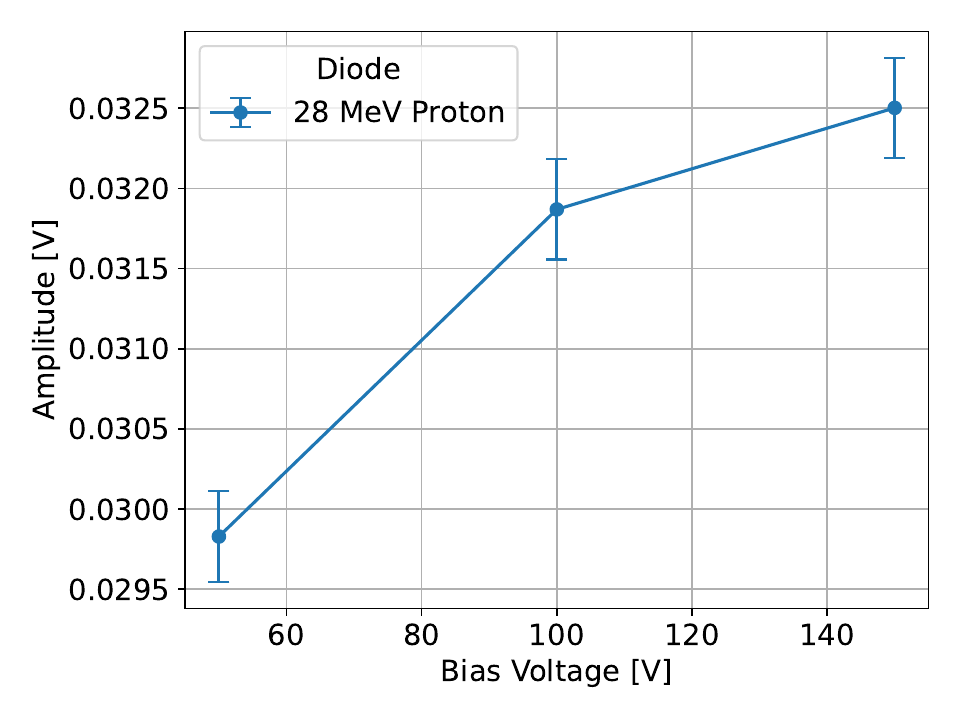}%{Picture1}
         \caption{}
         \label{fig:Diode_amplitudeVsVoltage}
     \end{subfigure}
     \hfill
          \begin{subfigure}[t]{0.47\textwidth}
         \centering
         \includegraphics{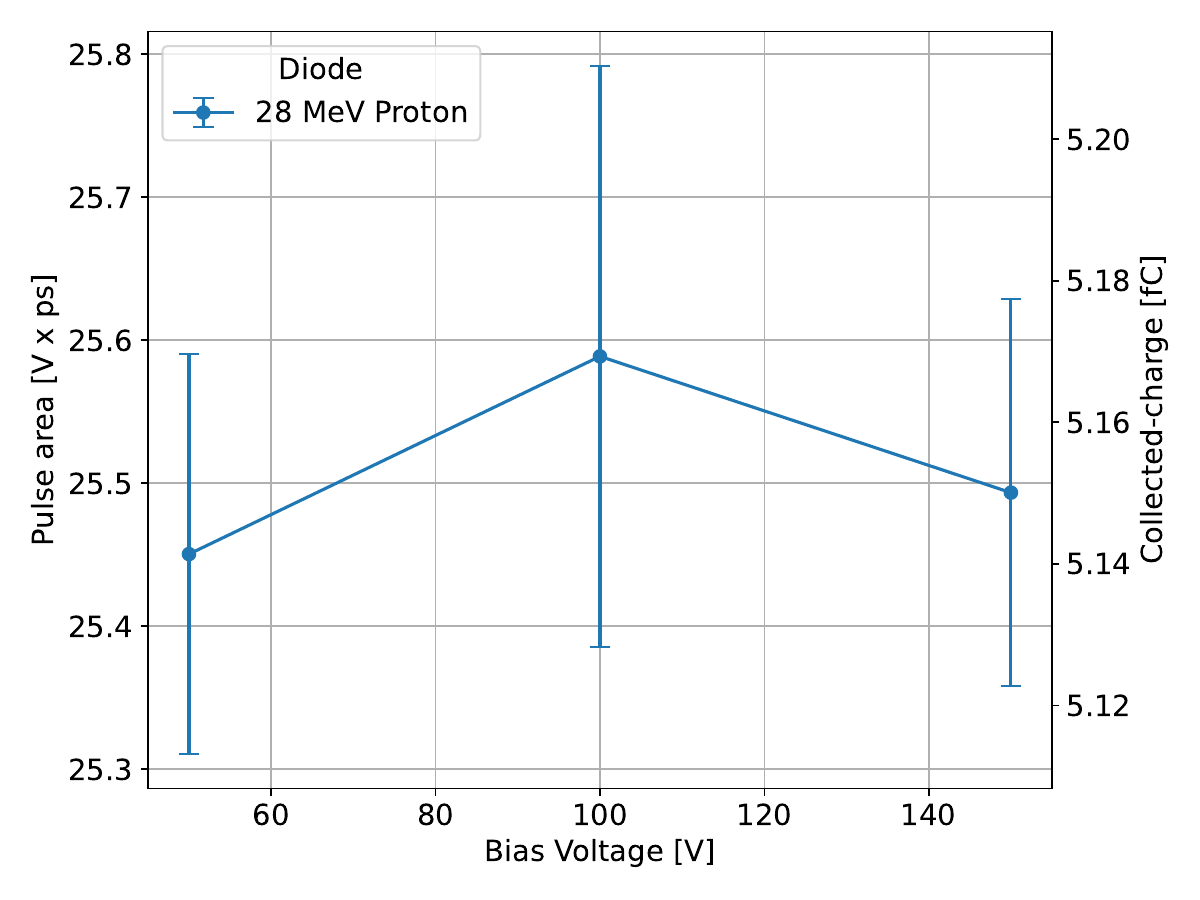}%{Picture3}
         \caption{}
         \label{fig:Diode_AreaVsVoltageVsVoltage}
         \end{subfigure}
     \hfill
     \begin{subfigure}[t]{0.45\textwidth}
         \centering
         \includegraphics{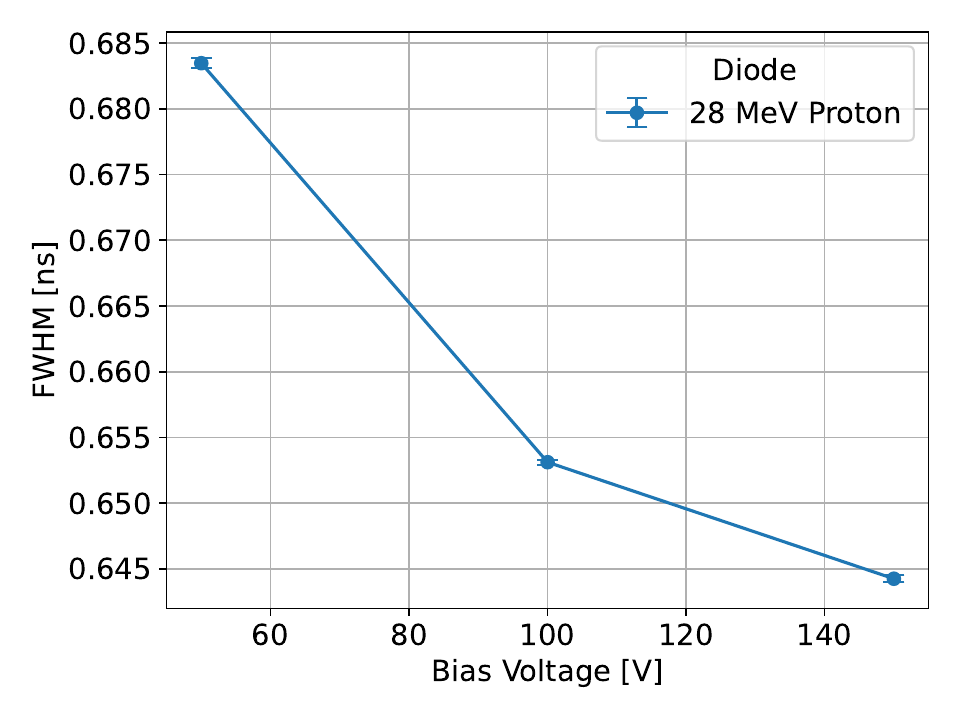}%{Picture2}
         \caption{}
         \label{fig:Diode_FHWMVsVoltage}
     \end{subfigure}
     \hfill
     \begin{subfigure}[t]{0.45\textwidth}
         \centering
         \includegraphics{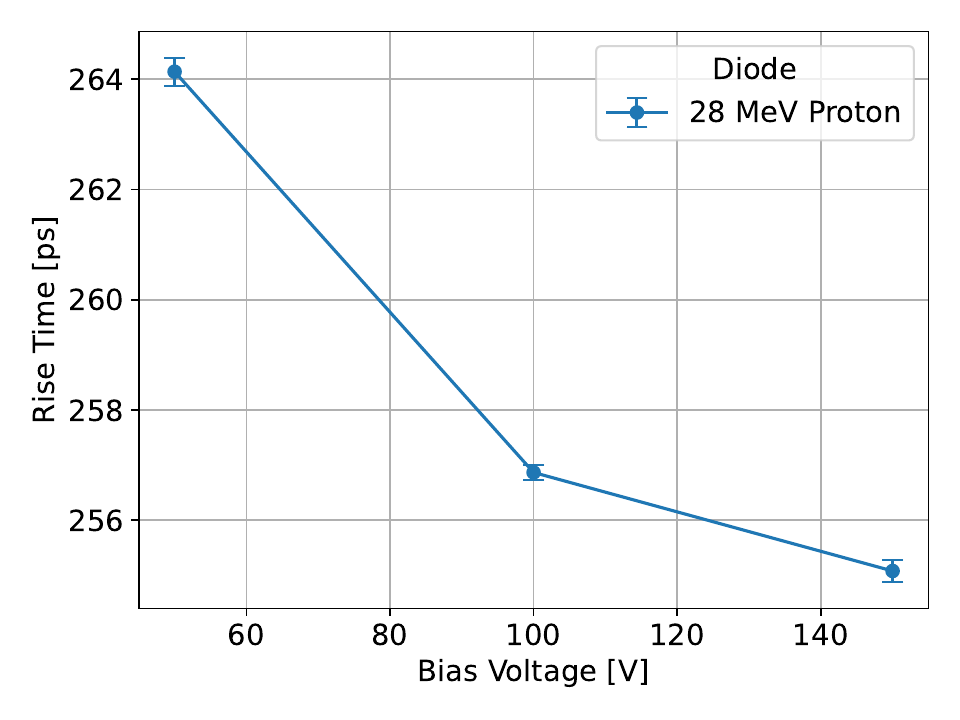}%{Picture3}
         \caption{}
         \label{fig:Diode_RiseTimeVsVoltage}
     \end{subfigure}
     \caption{Most probable values of amplitude (a), pulse area (b), FWHM (c) and rise time (d) as functions of the bias voltage with the 28-MeV proton beam for Diode-1. The most probable values and their uncertainties are extracted from Gaussian fits in limited
ranges around the peaks of the distributions.}
    \label{fig:signal-prop_diode_28MeVproton_Vbias}
\end{figure}

Figure~\ref{fig:signal-prop_diode_28MeVproton} and ~\ref{fig:signal-prop_diode_28MeVproton_Vbias} show amplitude, area, FWHM, and rise time as functions of bias voltage for Diode-1 exposed to a 28-MeV proton beam. The amplitude and area distributions increase little as the bias voltage increases, whereas the FWHM and rise time distributions show significant dependence on the bias voltage: the pulses become narrower and faster as the bias voltage increases.
%%%
%%%%
% \begin{figure}[bt]
% %\centering
% \includegraphics[width=.45\textwidth]{figures/Amplitude_combined_Diode.pdf}
% \includegraphics[width=.45\textwidth]{figures/Area_combined_Diode.pdf}
% \includegraphics[width=.45\textwidth]{figures/FWHM_combined_Diode.pdf}
% \includegraphics[width=.45\textwidth]{figures/RiseTime_combined_Diode.pdf}
% \caption{Most probable values of amplitudes (left) and pulse areas (right) as functions of the bias voltage with the 28-MeV proton beam for Diode-1. The most probable values and their uncertainties are extracted from Gaussian fits in limited
% ranges around the peaks of the distributions.}
% \label{fig:HistogramAmplitudeAndArea_Diode1}
% \end{figure}

%%%---------------------------
\subsection{Signal properties in LGADs}
\label{sec:signal_lgads}

\begin{figure}[t]
%\centering
\includegraphics[width=.5\textwidth]{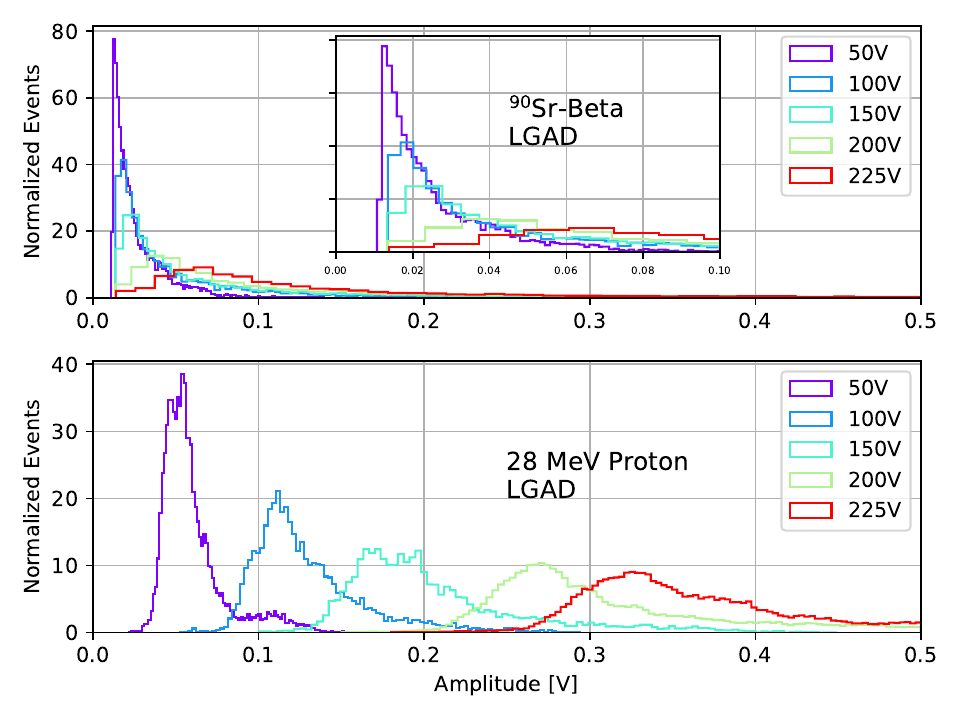}
\includegraphics[width=.5\textwidth]{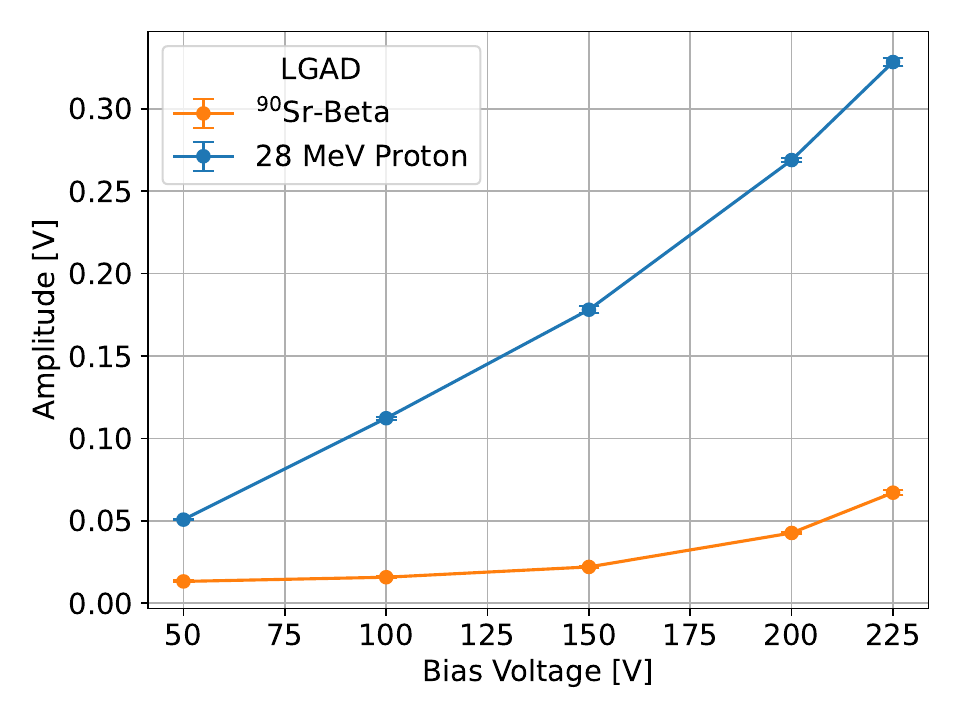}
\caption{Area-normalized distributions of amplitudes (left)  as functions of bias voltage in the LGAD, for beta-particles (left-top) and 28-MeV protons (left-bottom). Most probable values of amplitudes (right) as functions of bias voltage with beta-particles and 28-MeV protons, for the LGAD. The most probable values and their uncertainties are extracted from Gaussian fits in limited ranges around the peaks of the distributions shown on the left.}
\label{fig:HistogramAmplitudeLGADs}
\end{figure}

Unlike the diodes, the LGAD shows strong dependence of pulse amplitude, area, FWHM, and rise time on bias voltage.
Figure~\ref{fig:HistogramAmplitudeLGADs} shows the amplitude distributions in the LGAD when it is exposed to 28-MeV protons and beta-particles at different bias voltages. For both particle beams the amplitudes increase with increasing values of the bias voltage. The amplitudes are larger in a 28-MeV proton beam than in a beta-particle beam. At a bias voltage of 50 V, for the beta-particle beam, the lower-end of the amplitude distribution is cut by the oscilloscope trigger, therefore such a data set is not used in the following analyses. At a bias voltage of 225 V, the most probable value of the amplitude distribution reaches 67 mV with beta-particles and 328 mV with 28-MeV protons. %Figure~\ref{fig:AmplitudeMeanValues} confirms the above results.

A double peaked structure was observed in the LGAD amplitude distributions; a smaller bump at low amplitudes, compatible with the amplitudes observed in diodes, was seen in addition to the main peak. This is an effect of the limited area of the gain layer with respect to the whole sensitive area, which includes the junction termination extensions and guard-rings: particles that impinge upon the LGAD at the edges of the sensitive area, outside the $p^+$ layer, are not multiplied, and deformations in the electric fields make them drift  towards and be collected by the edge of the pad, where the electric field is low, i.e.\ there is no gain.  Such events are removed from the event sample for further analysis by analyzing the distribution of the amplitude as a function of the FWHM, which shows a clear demarcation region between the two sets of events.

\begin{figure}[bt]
%\centering
\includegraphics[width=.55\textwidth]{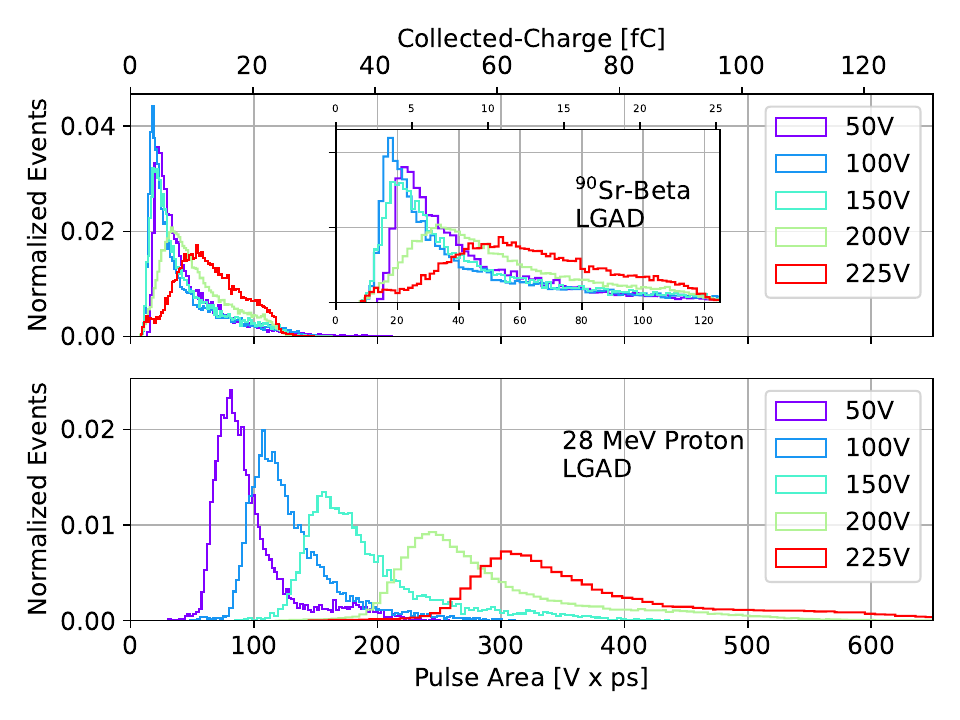}
\includegraphics[width=.50\textwidth]{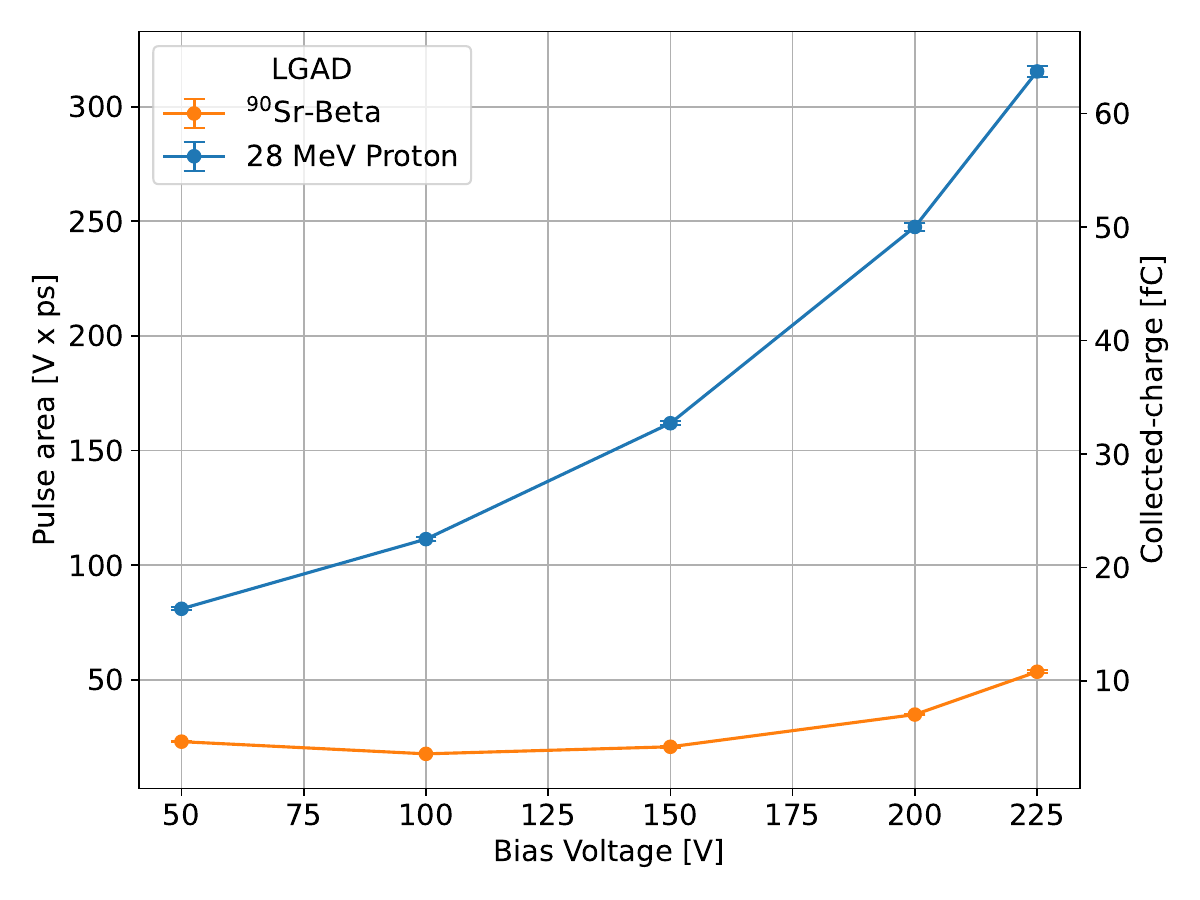}
\caption{Area-normalized distributions of pulse area (left) as functions of bias voltage in the LGAD, for a beam of beta-particles (left-top) and 28-MeV protons (left-bottom). Most probable values of pulse areas (right) as functions of bias voltage with beta-particle and 28-MeV proton beams for the LGAD. The most probable values and their uncertainties are extracted from Gaussian fits in limited ranges around the peaks of the distributions shown on the left.}
\label{fig:HistogramAreaLGADs}
\end{figure}

The distributions of pulse areas in the LGAD, shown in Figure~\ref{fig:HistogramAreaLGADs}, have a trend very similar to that of the distribution of amplitudes for both beam types, as a function of the bias voltage.

%%%

\begin{figure}[bt]
%\centering
\includegraphics[width=.5\textwidth]{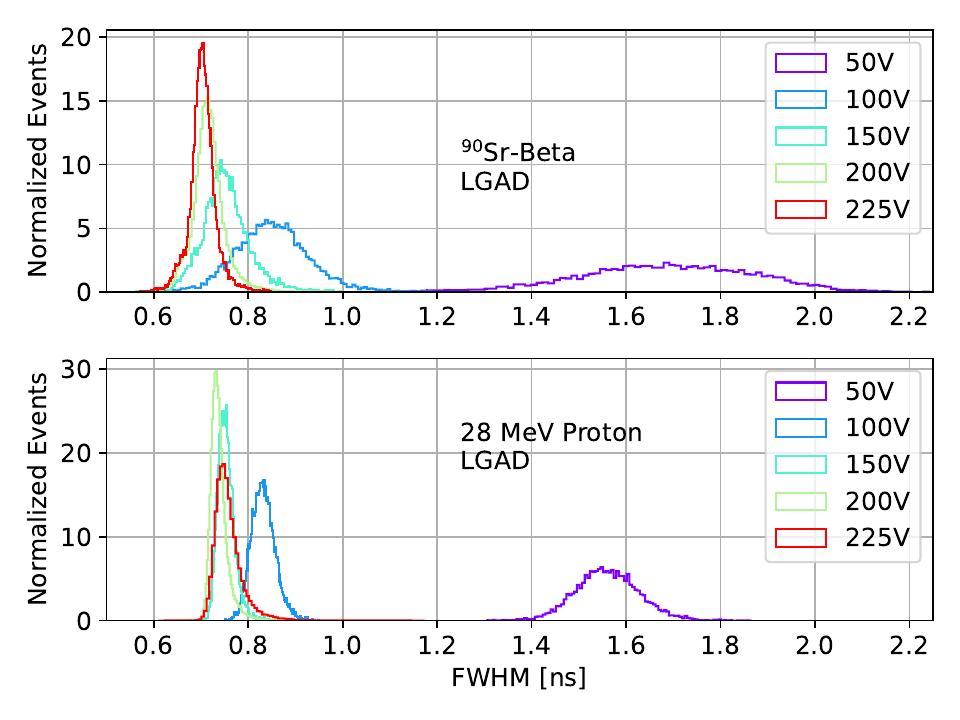}
\includegraphics[width=.5\textwidth]{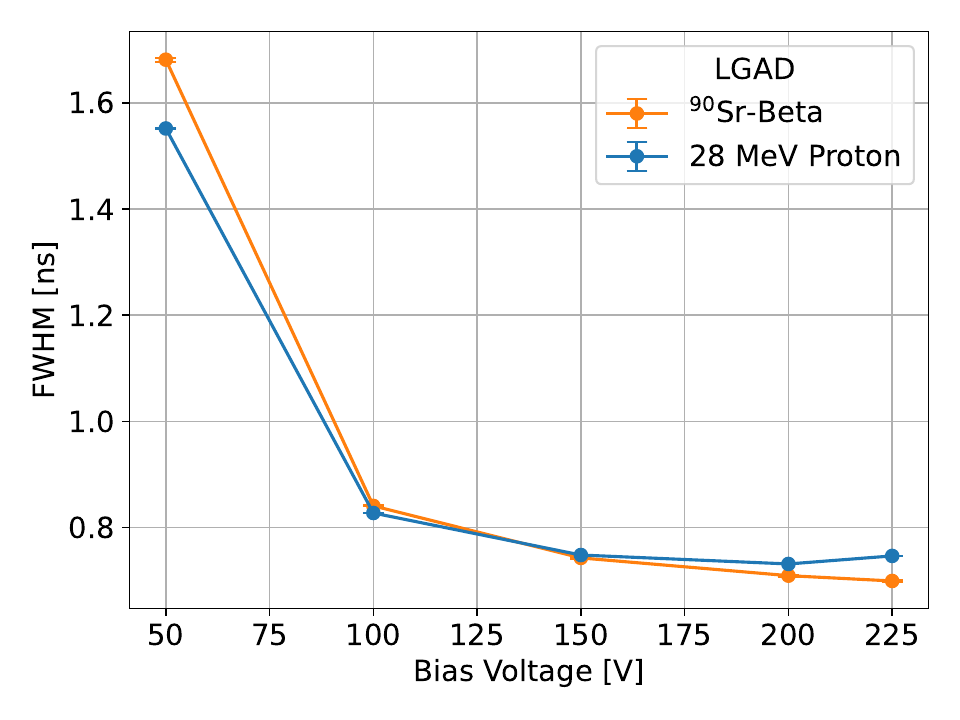}
\caption{Area-normalized distributions of FWHM (left) as functions of bias voltage in the LGAD, for a beam of beta-particles (left-top) and 28-MeV protons (left-bottom). Most probable values of FWHM (right) as functions of bias voltage with  beta-particle and 28-MeV proton beams, for the LGAD. The most probable values and their uncertainties are extracted from Gaussian fits in limited ranges around the peaks of the distributions shown on the left.}
\label{fig:HistogramFWHMLGADs}
\end{figure}

At low bias voltage, i.e.\ at 50 V, the LGAD pulse is very broad for both beam types, with a FWHM of about 1.5-1.9 ns, see Figure~\ref{fig:HistogramFWHMLGADs}. The pulses become narrower when the LGAD is biased from 100 V upwards, reaching a FHWM as small as about 700-750 ps with beta-particles. The LGAD shows very similar values of FWHM with beta-particles and 28-MeV protons.    %Figure~\ref{fig:FWHMMeanValues} confirms the above observations.
%and shows more clearly that with 28-MeV protons LGAD-2 has far broader pulses, with FWHM values that are up to a factor of two greater than those in LGAD-1, when the sensors are fully depleted. 

\begin{figure}[ht]
%\centering
\includegraphics[width=.5\textwidth]{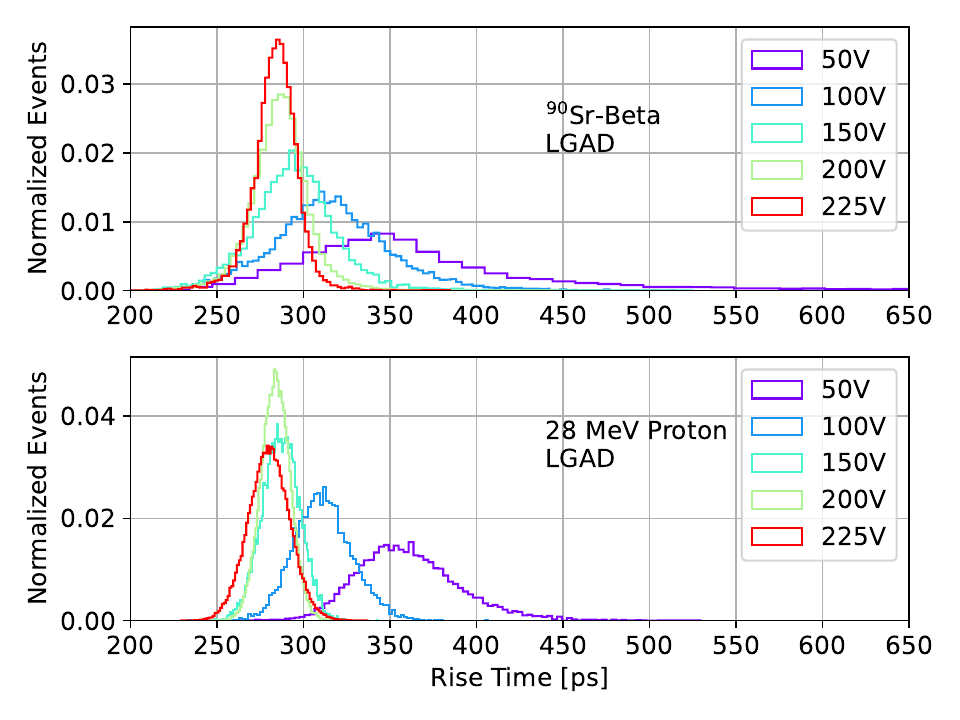}
\includegraphics[width=.5\textwidth]{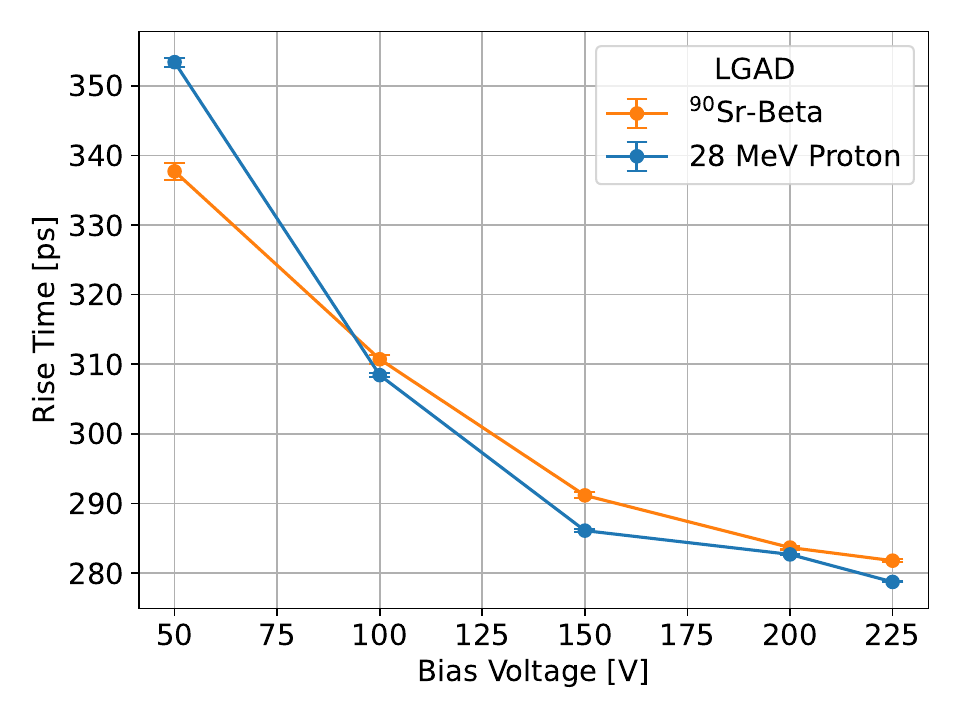}
\caption{Area-normalized distributions of rise time (left) as functions of bias voltage in the LGAD, for a beam of beta-particles (left-top) and 28-MeV protons (left-bottom). Most probable values of rise time (right) as functions of bias voltage with beta-particle and 28-MeV proton beams for the LGAD. The most probable values and their uncertainties are extracted from Gaussian fits in limited ranges around the peaks of the distributions shown on the left.}
\label{fig:HistogramRiseTimeLGADs}
\end{figure}

The rise times in Figure~\ref{fig:HistogramRiseTimeLGADs} show that the LGAD signal becomes faster as the bias voltage increases. More specifically, the peak of the rise time distribution decreases from about 340 ps at a bias voltage of 50 V to about 280 ps at a bias voltage of 225 V, for a beta-particle beam. The rise time distributions are very similar between the two beam types at each value of bias voltage for the LGAD. 
%For example, at 225 V bias voltage, LGAD-2 has a rise time close to 100 ps larger than LGAD-1.

\section{Gain Measurements} 
\label{sec:gain}
%%%%%%%%%%%%%%%%%%%%%%%%%%%%%%%%%%%%%%%%%%%%%%%%%

The gain is measured for the LGAD using two different methods that are often used to estimate the gain. The gain is measured as the ratio of the most probable signal amplitudes (first method) or collected charges (second method) in the LGAD and a diode under the same experimental conditions.  
%The first method assumes a proportionality between the generated charge and the signal amplitude, while the second method assumes a proportionality between the generated charge and the measured charge, in a sensor. 
Since different assumptions and experimental techniques are used in both methods, both sets of results are presented.

%%%%%%%%%%%%%%%%%%%%%%%%%
%%%----------------------
\subsection{Gain from signal amplitudes}
\label{sec:gain_ampl}

In this method the most probable values of the amplitudes are extracted for the LGAD in both beam types and for diodes in the 28-MeV proton beam from a Gaussian fit in a restricted range of the amplitude distribution around the peak. The fit range was chosen to minimize the $\chi^2$/n.d.f.\ and was not greater than the full width at half maximum of the amplitude distribution.
%The fit range is optimized base on the quality of the fit, based on a $\chi^2$ method. 
The fits result in $\chi^2$/n.d.f.\ values typically $\le 1$ with the exception of about four amplitude distributions for the case of beta-particle beams, which have values between 1 and 5. For the few cases in which the $\chi^2$/n.d.f.\ significantly exceeded 1, the uncertainties were adjusted to reflect the poor fit quality.

%%%-----------------------
\subsubsection{Gain with 28-MeV protons}
\label{sec:gain_ampl_p}

\begin{figure}[t]
% \centering
\includegraphics[width=.5\textwidth]{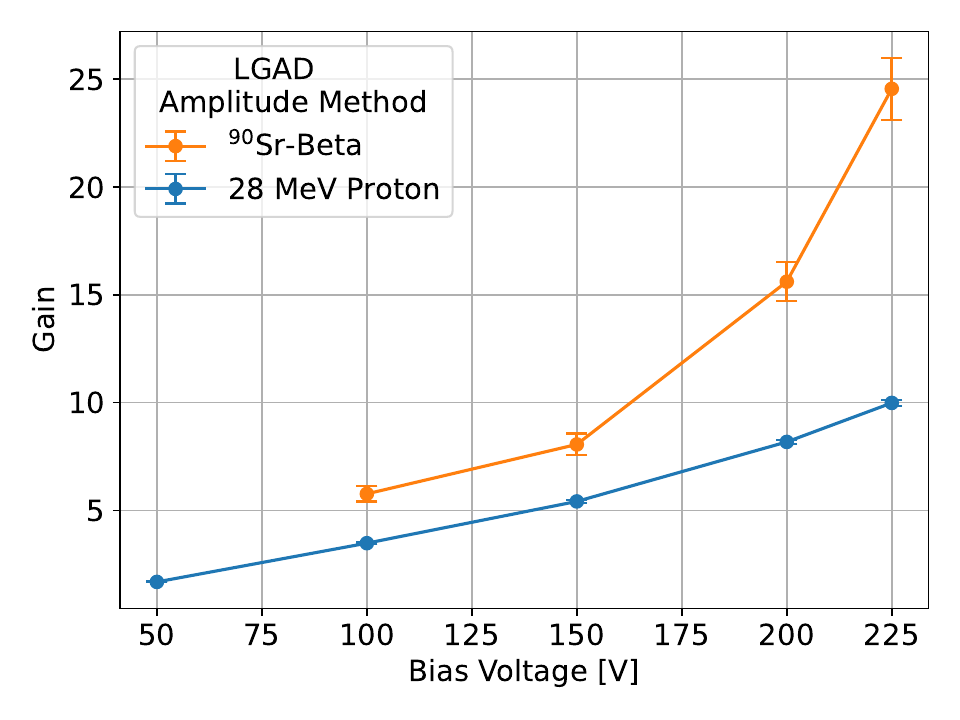}
\includegraphics[width=.5\textwidth]{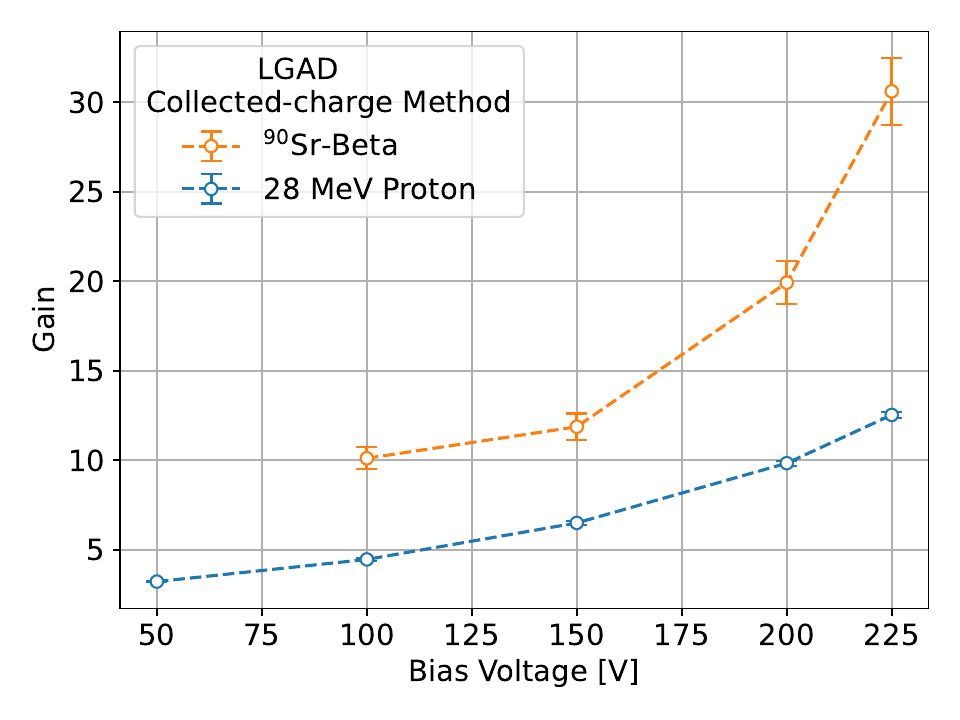}
\caption{Gain of the LGAD as a function of bias voltage in a 28-MeV proton beam and with beta particles, using the amplitude method (Left) and the collected-charge method (Right). The data points at 50 V bias voltage are not shown for the beta-particle beam, because the associated pulse amplitudes and areas in data are truncated due to the triggering threshold in the oscilloscope (see Figures~\ref{fig:HistogramAmplitudeLGADs},~\ref{fig:HistogramAreaLGADs}).
}
\label{Gain_LGAD_Amplitudes}
\end{figure}

In the 28-MeV proton beam, the gain is measured as the ratio of the most probable signal amplitude in the LGAD, as numerator, and the most probable signal amplitude in a diode, as denominator. The gain is measured as a function of the bias voltage for the LGAD and a diode up to 150~V. For bias voltages above 150 V, the amplitudes in the diode are taken as the values measured at 150 V, while those in the LGAD are taken at the corresponding voltage, i.e.\ 200 V and 225 V. 
%\textcolor{orange}{Hijas: gain is measured by dividing the LGAD amplitude for particular voltage(i.e 50V)/Diode Amplitude for that particular voltage(i.e 50V) till 150V, after 150V, for 200V and 225V, Diode parameter is taken as the value corresponding to 150V (The highest Diode Voltage we took data for)} 
The gain is measured as the average of the two gains obtained with the two diodes as denominators. The relative uncertainties from the fits of the amplitude distributions for the LGAD and diodes are included in the gain measurements. Figure~\ref{Gain_LGAD_Amplitudes} shows the dependence of the gain of the LGAD on the bias voltage. %They show a similar trend but LGAD-2 has a lower gain than LGAD-1: 
The gain of the LGAD ranges from about 1.6 at 50 V to 10 at 225 V.
%, while the gain of LGAD-2 is in the range 1.2 - 6 in the 50 - 225 V bias-voltage range.

%%%
%%%
% \begin{figure}[t]
% \centering
% \includegraphics[width=.5\textwidth]{figures/28MeVProton_gainSummaryAmplitude.png}
% \caption{Gains of LGAD-1 and LGAD-2 as functions of bias voltage in a 28-MeV proton beam, using the amplitude method. %The uncertainties include those from the amplitude distribution fits and from the use of the two Diodes in the calculation, as explained in the text.
% }
% \label{Gain_28MeVProton_Amplitudes}
% \end{figure}
% AT:  The uncertainy on L is added linearly, whule the unceryainties on d1 and d2 are summed quadratically to each other as L is correlated between g1 and g2 measurements whereas d1 and d2 are uncorrelated 
%\textcolor{orange}{Hijas:\\
%gain error estimation
%\begin{equation}
%    g1 = \frac{LGAD (L)}{Diode1 (d1)}
%\end{equation}
%\begin{equation}
%    g2 = \frac{LGAD (L)}{Diode2 (d2)}
%\end{equation}
%\begin{equation}
%    \Delta g = \Delta L +  \sqrt{(g1(\frac{\Delta d1}{ d1})^2+(g2(\frac{\Delta d2}{d2})^2}
%\end{equation}
%Error ($\Delta$ ) is the sigma of Gaussian fit of Amplitude/Area distribution\\
%Un-correlated errors were calculated this way
%}
%%%-----------------------
\subsubsection{Gain with beta-particles}
\label{sec:gain_ampl_beta}

Since the most probable value of the amplitude distribution cannot be accurately measured with beta-particles in diodes, it is extrapolated using the following method.

\begin{figure}[tb]
\centering
\includegraphics[width=.45\textwidth]{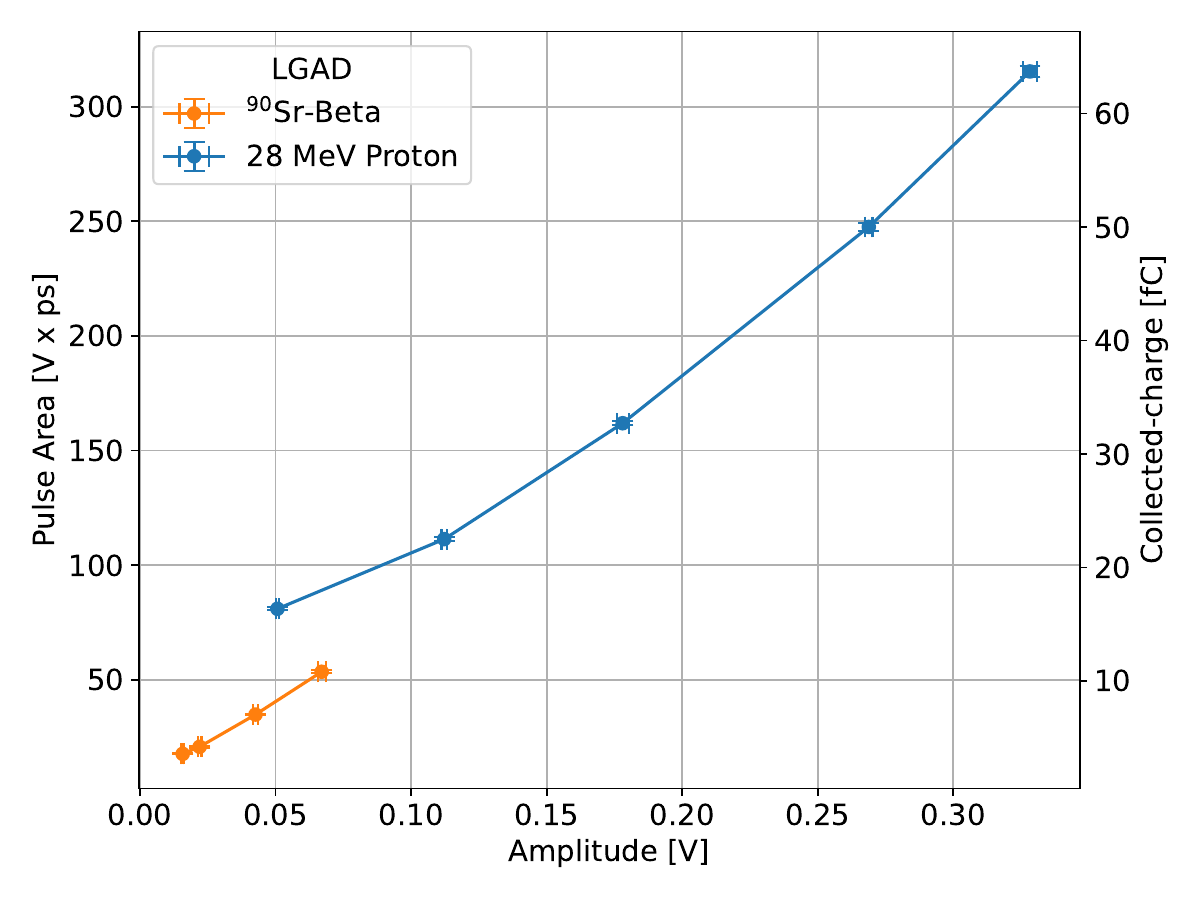}
\includegraphics[width=.45\textwidth]{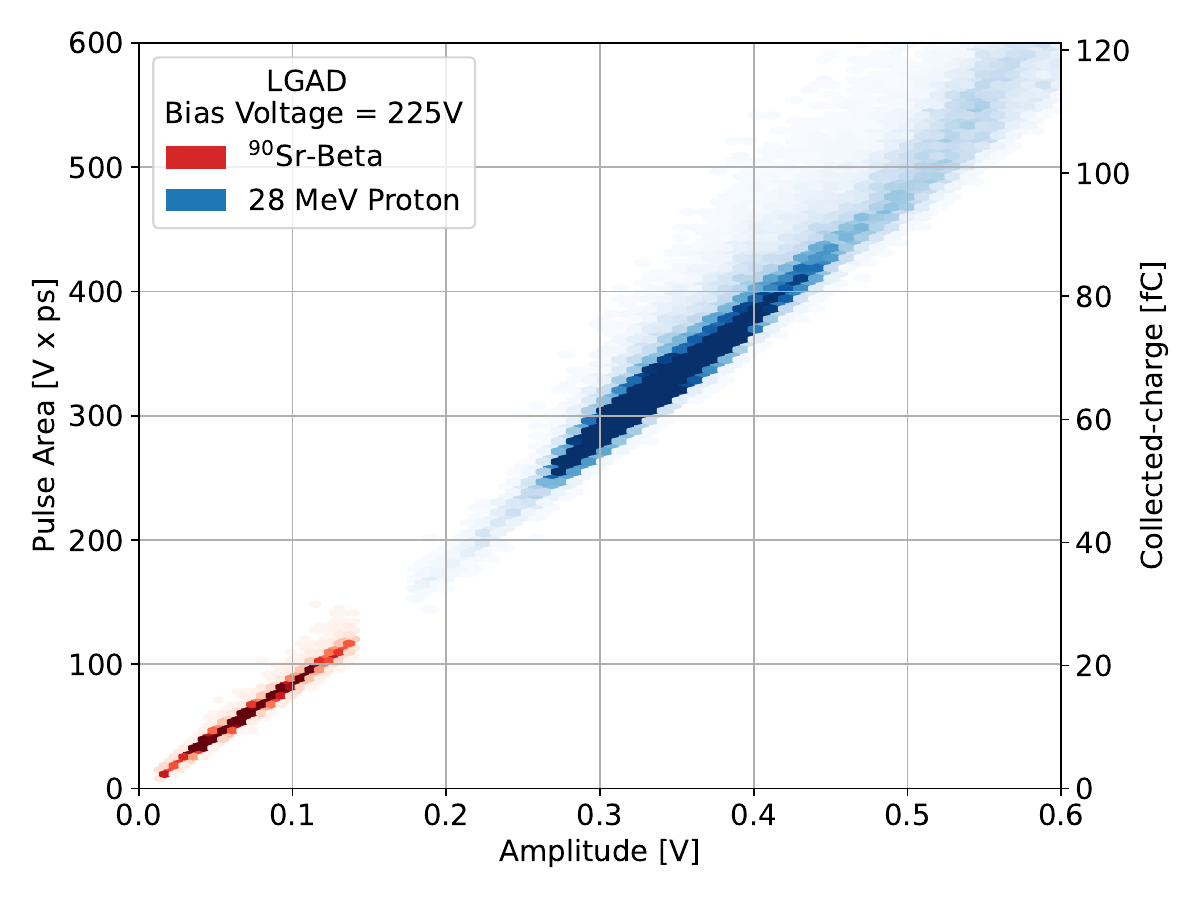}
\caption{(Left) Mean values of the pulse area as a function of the mean values of the pulse amplitude of the LGAD in a 28-MeV proton beam and with beta-particles at the five considered bias voltages, i.e.\ 100, 150, 200 and 225 V. The data points at 50 V bias voltage are not shown for the beta-particle beams, because the associated pulse amplitudes and areas in data are truncated due to the triggering threshold in the oscilloscope (see Figures~\ref{fig:HistogramAmplitudeLGADs},~\ref{fig:HistogramAreaLGADs}). (Right) Areas as a function of amplitudes measured on a pulse-by-pulse basis for the samples collected at a bias voltage of 225 V for both beam types.  
}
\label{fig:AreaVsAmplitudes}
\end{figure}

The measured collected charge in a diode with a MIP as incident particle  corresponds to the integral of a pulse (Area) and can be expressed as follows:
\begin{equation}
    {\rm Area} = \int dt V =  Q R A^2 .
\end{equation}
Here $V$ is the signal voltage sampled in a time range $dt$; $Q$ is the charge generated in 35 $\mu$m thick silicon, which is computed using $62.9 \pm 3.8$ as the number of electron-hole pairs generated per micron by a MIP (see Appendix~\ref{app:n_ehPairs_meas}); $R$ is the amplifier resistance, $50~\Omega$; and $A$ is the amplifier gain in the FNAL board, which is measured to be $9.95\pm 0.02$, using data collected with a beam of X-rays from an $^{241}$Am source (see Appendix~\ref{app:amplifier_A}). Thus, the value of the Area for a MIP is calculated to be (1.75 $\pm$ 0.11) V ps.
The Area was experimentally observed to be linearly proportional to the amplitude, as shown in Figure~\ref{fig:AreaVsAmplitudes}. 
%%%
%%%
Thus, the most probable value of the amplitude for a MIP is obtained by extrapolating a linear fit to data from beta-particles in the two-dimensional area-amplitude distribution to the Area value of (1.75 $\pm$ 0.11) V ps (as calculated above). This results in an amplitude of $(2.73\pm 0.15)$ mV for a diode biased at 100 V. A compatible value is found at higher bias voltages. 

The uncertainty on the gain includes the quadrature sum of the relative uncertainty on gain of the electronic board and that on the measured number of electron-hole pairs generated in a micron in silicon. The total uncertainty is dominated by those on the linear fit parameters. Systematic variations of the parametric function used in the fit, i.e.\ polynomial of different degrees and exponential, have a negligible impact on the diode's amplitude uncertainty: the uncertainty of the linear fit is 5.45\% while the largest difference between linear fit and other parametrization is 0.02\%.

Figure~\ref{Gain_LGAD_Amplitudes} shows that the gains with the amplitude method in a beam of beta-particles are in the approximately 5-25 range for bias voltages in the 100-225 V range, and they increase as the bias voltage increases. 

%%%%%%%%%%%
\begin{figure}[tb]
% \centering
      \setkeys{Gin}{width=\linewidth}
     \begin{subfigure}[t]{0.5\textwidth}
      \centering
         \includegraphics{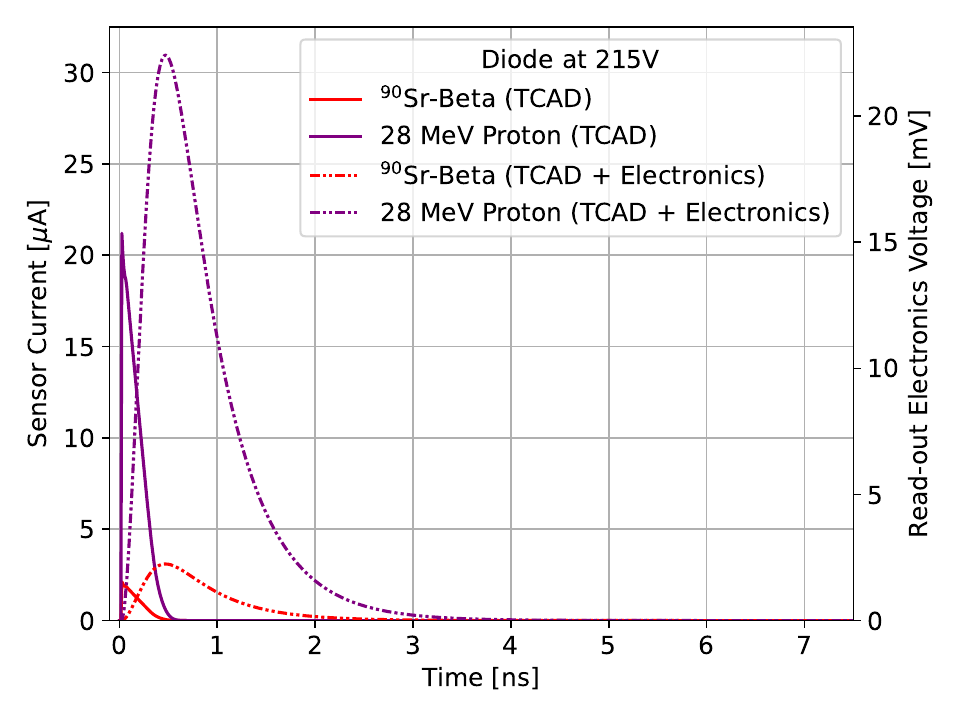}%{Picture1}
         \caption{Diode Simulation}
         \label{fig:waveform_simulation_diode}
     \end{subfigure}
     \hfill
     \begin{subfigure}[t]{0.5\textwidth}
         \centering
         \includegraphics{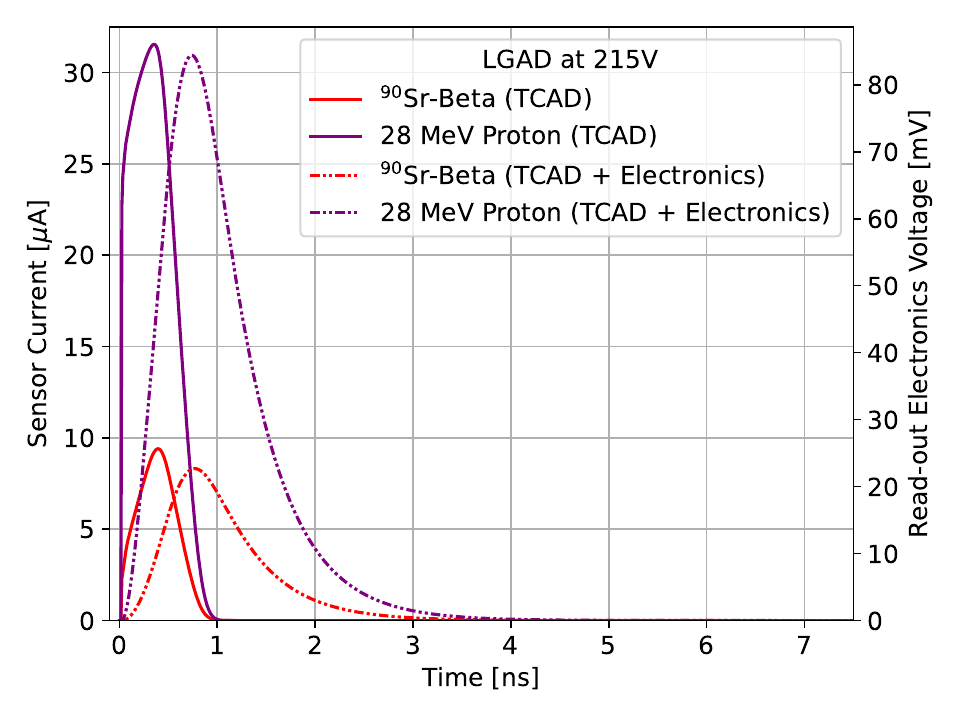}%{Picture2}
         \caption{LGAD Simulation}
         \label{fig:waveform_simulation_lgad}
     \end{subfigure}
     \caption{TCAD simulation and effect of electronics on simulation for pulses generated in a diode (left) and an LGAD (right) by a 28-MeV proton and a beta-particle from a $^{90}{\rm Sr}$ radioactive source. In all cases, the bias voltage is set to 215 V.}
    \label{fig:Waveform_simulations}
\end{figure}
%\FloatBarrier

%%%%%%%%%%%%%%%%%%%%%%%%%
%%%----------------------
\subsection{Gain from collected charge}
\label{sec:gain_charge}

In this method the collected charges are measured for the LGAD in both beam types and for diodes in the 28-MeV proton beam by using the signal pulse area. The distributions of the pulse areas are then fit using a Gaussian function in a limited range that was chosen to minimize the $\chi^2$/n.d.f.\ and was not greater than the full width at half maximum of the distribution. The fits result in $\chi^2$/n.d.f.\ values generally $\le 1$.
%\textcolor{orange}{Hijas: Integral is done by taking the area under the signal pulse, where signal pulse is divided into smaller trapeziums of height 50 ps (time interval of oscilloscope voltage measurements), and the length of the parallel lines are the height of the signal pulse in V for adjacent readings, area of each trapezium is summed to get the integral of the signal pulse}

%%%-----------------------
\subsubsection{Gain with 28-MeV protons}
\label{sec:gain_charge_p}

The procedure for the measurement of the gain with 28-MeV protons using the collected-charge method is similar to the one explained in Section~\ref{sec:gain_ampl_p} for the amplitude method. In this method, the gain is measured as the ratio of the most probable pulse area value in the LGAD, as numerator, and the most probable pulse area value in a diode, as denominator. Figure~\ref{Gain_LGAD_Amplitudes} shows the gain for the LGAD as a function of the bias voltage. The uncertainties are estimated using the same method as in Section~\ref{sec:gain_ampl_p}. In this method the LGAD gains are in the range approximately 3-13.

%%%%%%%%%%%%%%%%%%%%%%%%%%%%%%%%%%%%%%%%%%%%%%%
\subsubsection{Gain with beta particles}
\label{sec:gain_charge_beta}

For the estimate of the collected charge in a diode with a beta-particle beam, the same calculations as are described in Section~\ref{sec:gain_ampl_beta} are used. Thus, a pulse area value of 1.75 V ps is used for a MIP as the denominator in the gain calculation. Figure~\ref{Gain_LGAD_Amplitudes} shows the dependence of the gain as a function of the bias voltage.

%%%%%%%%%%%%%%%%%%%%%%%%%%%%%%%%%%%%%%%%%%%%%%%%%
\section{Simulation}
\label{sec:gain_sim}
%%%-----------------------
%%%%%%%%%%
\begin{figure}[tb]
% \centering
      \setkeys{Gin}{width=\linewidth}
     \begin{subfigure}[t]{0.42\textwidth}
         \includegraphics[width=1.15\textwidth]{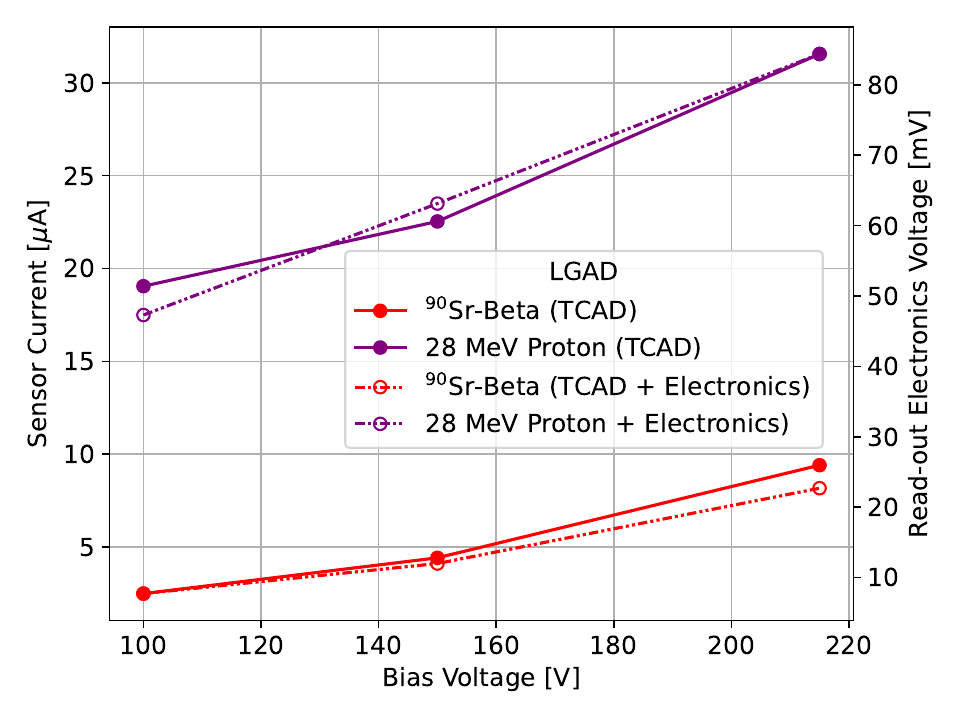}%{Picture1}
         \caption{Simulated Amplitude}
         \label{fig:AmplitudeSimul}
     \end{subfigure}
     \hfill
     \begin{subfigure}[t]{0.42\textwidth}
         %\centering
         \hspace{-1cm}
         \includegraphics[width=1.15\textwidth]{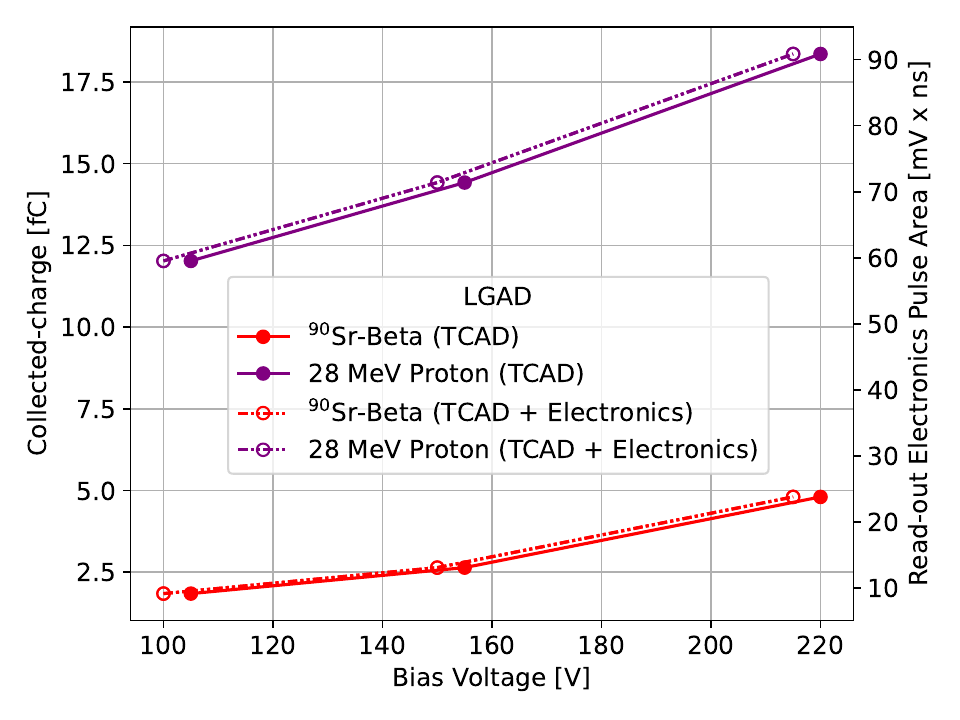}%{Picture3}
         \caption{Simulated Area}
         \label{fig:RiseTimeSimul}
     \end{subfigure}
     \hfill
     \begin{subfigure}[t]{0.38\textwidth}
         %\centering
         \includegraphics[width=1.15\textwidth]{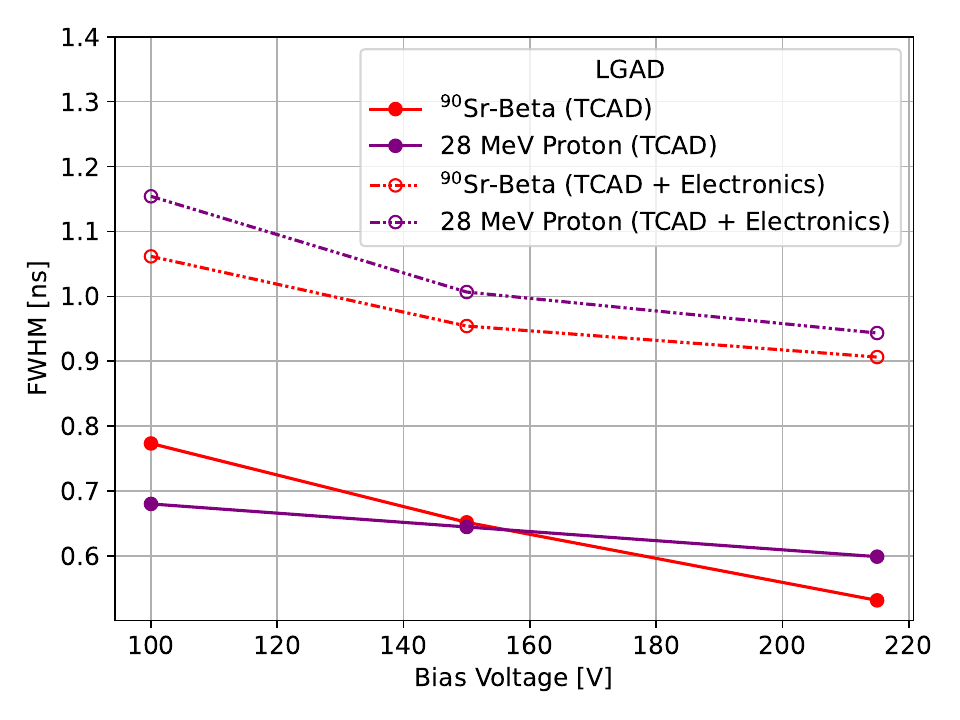}%{Picture2}
         \caption{Simulated FWHM}
         \label{fig:FWHMSimul}
     \end{subfigure}
     \hfill
     \begin{subfigure}[t]{0.38\textwidth}
         %\centering
         \hspace{-1.4cm}
         \includegraphics[width=1.15\textwidth]{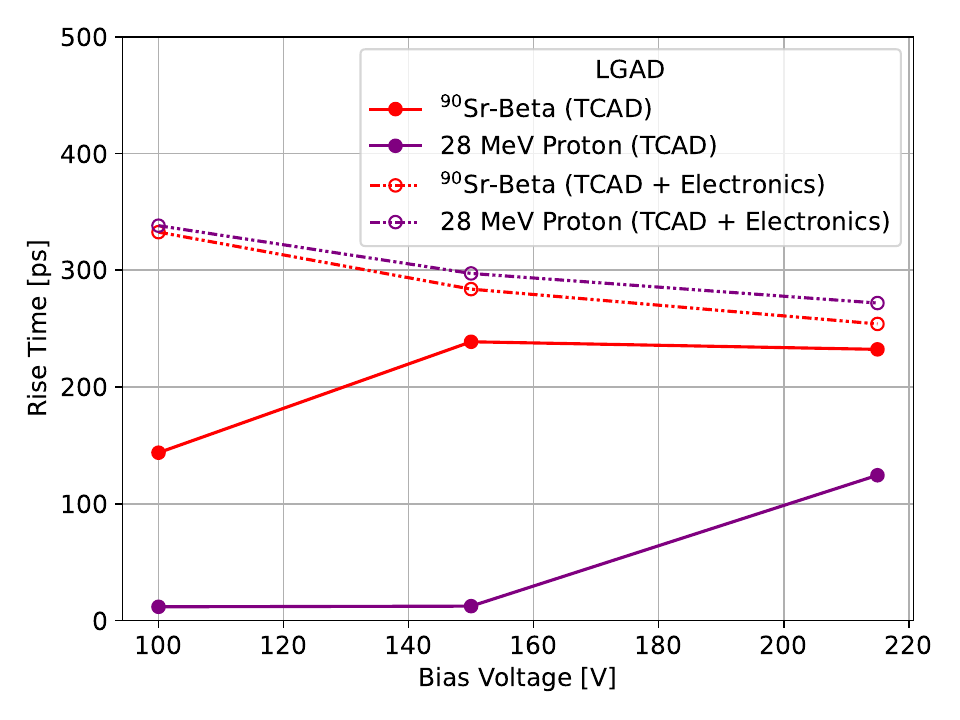}%{Picture3}
         \caption{Simulated Rise time}
         \label{fig:RiseTimeSimul}
     \end{subfigure}
     \caption{Simulated amplitude (a), area (b), FWHM (c) and rise time (d)  in beta-particles and 28-MeV protons as functions of bias voltage, for an LGAD, with and without the effect of the readout electronics. In (b) the lines for the TCAD simulation and the one including the effect of the electronics are very close, therefore the markers of the former have been slightly shifted along the $x$-axis for display purposes.
     }
    \label{fig:Amp-FWHM-Risetime-Charge_simulations_LGAD}
\end{figure}
%\FloatBarrier 
%%%%%%%%%%%%%
\begin{figure}[htb]
% \centering
      \setkeys{Gin}{width=\linewidth}
     \begin{subfigure}[t]{0.42\textwidth}
         \includegraphics[width=1.15\textwidth]{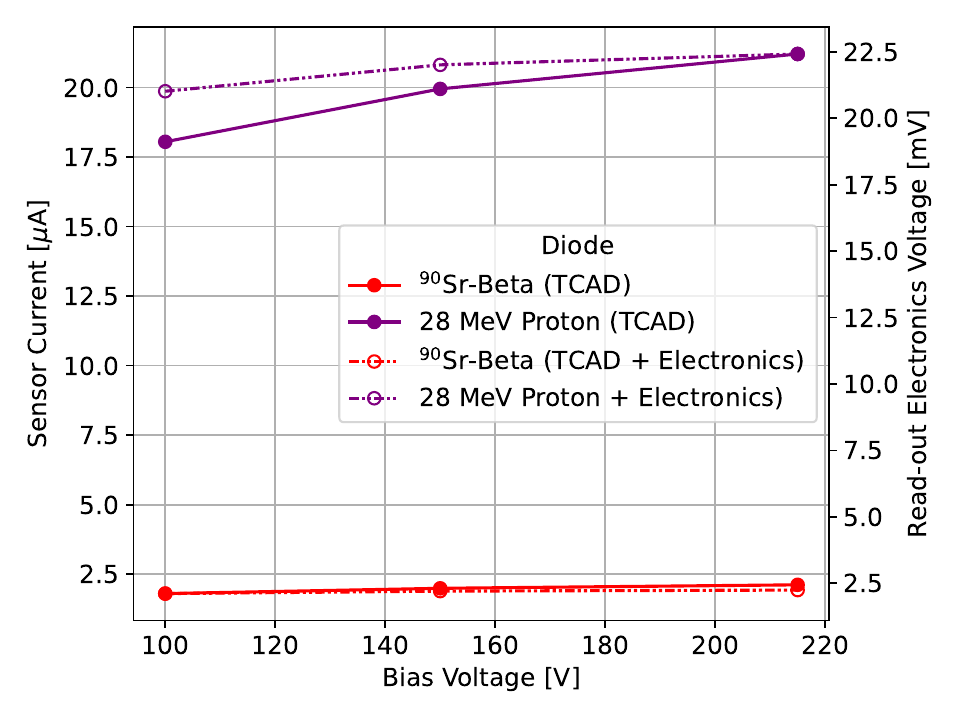}%{Picture1}
         \caption{Simulated Amplitudes}
         \label{fig:AmplitudeSimul}
     \end{subfigure}
     \hfill
     \begin{subfigure}[t]{0.42\textwidth}
         %\centering
          \hspace{-1cm}
         \includegraphics[width=1.15\textwidth]{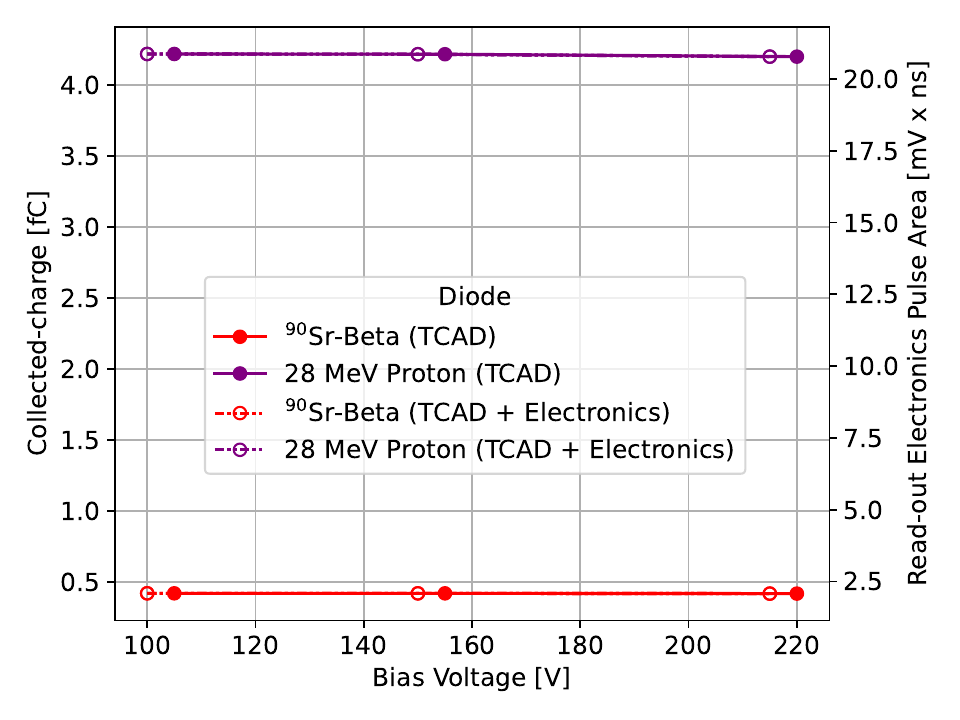}%{Picture3}
         \caption{Simulated Area}
         \label{fig:AreaSimul}
     \end{subfigure}
     \hfill
     \begin{subfigure}[t]{0.36\textwidth}
         %\centering
         \hspace{0.3cm}
         \includegraphics[width=1.15\textwidth]{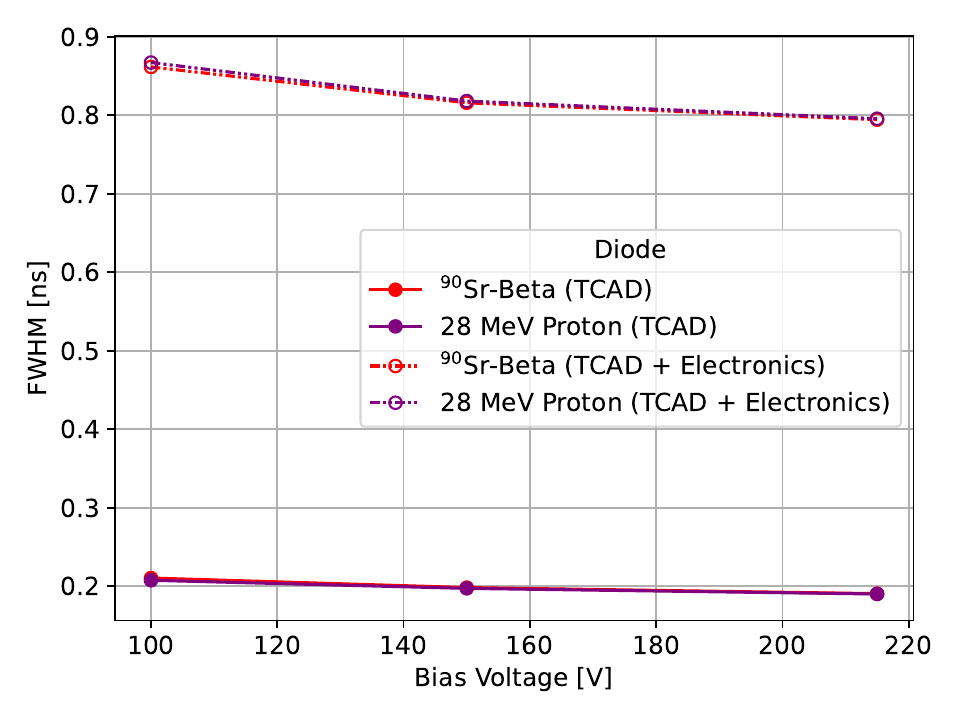}%{Picture2}
         \caption{Simulated FWHM}
         \label{fig:FWHMSimul}
     \end{subfigure}
     \hfill
     \begin{subfigure}[t]{0.36\textwidth}
         %\centering
        \hspace{-1.7cm}
         \includegraphics[width=1.15\textwidth]{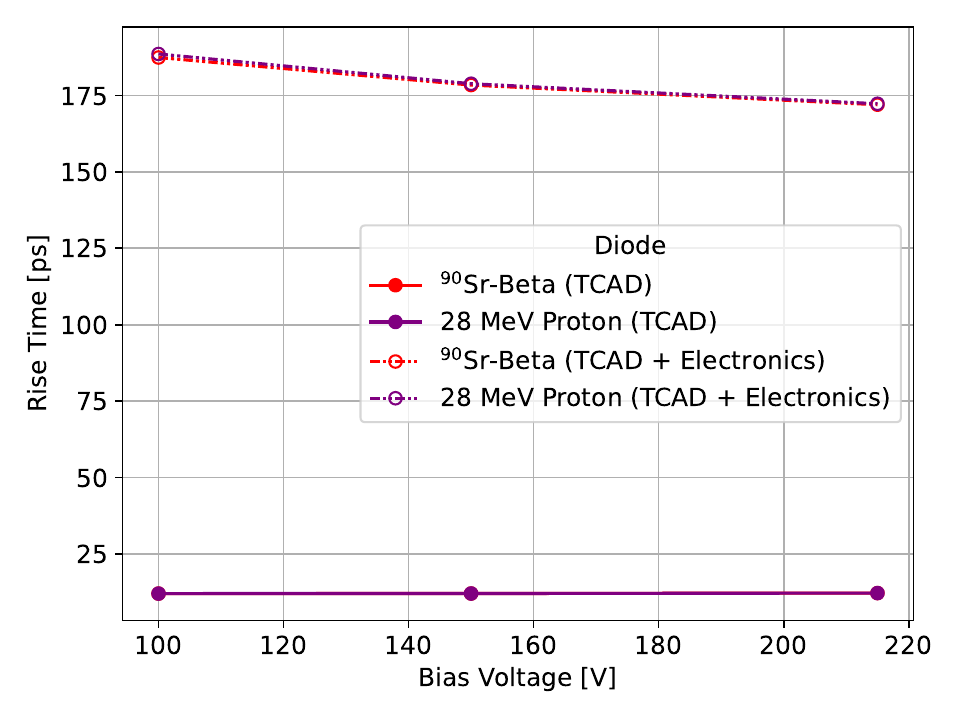}%{Picture3}
         \caption{Simulated Rise time}
         \label{fig:RiseTimeSimul}
     \end{subfigure}
     \caption{Simulated amplitude (a), area (b), FWHM (c) and rise time (d)  in beta-particles and 28-MeV protons as functions of bias voltage, for a diode with and without the effect of the readout electronics. In (b) the lines for the TCAD simulation and the one including the effect of the electronics overlap and the markers of the former have been slightly shifted along the $x$-axis for display purposes. In (c) and (d) the lines for the beta-particle beam and the 28-MeV proton beam overlap.  
     }
    \label{fig:Amp-FWHM-Risetime-Charge_simulations_Diode}
\end{figure}
%\FloatBarrier 
%%%%%%%%%%%%%%%

A SILVACO TCAD simulation was performed to predict the signal characteristics of a diode and an LGAD when they are exposed to beams of beta particles and 28-MeV protons. The geometrical characteristics and doping profiles of the sensors in simulation are qualitatively close to those of the devices under test. The simulation was carried out for the diode and the LGAD with the bias voltages was set at 100, 150 and 215 V. 

Figure~\ref{fig:Waveform_simulations} shows the predicted pulses for a diode and an LGAD as generated by the two different particle beams when the sensors are biased at 215 V. 
In Table~\ref{tab:params_simulations}, the amplitude, the pulse area, the FWHM and the rise times for the LGAD and the diode are extracted from the generated pulses with the bias voltage set at 215 V. The effects of the electronics on pulses are simulated using a first-order low-pass filter response function for the two amplification stages of the readout board and the one of the oscilloscope, using the readout board and oscilloscope parameters presented in Section~\ref{sec:DUT}.\footnote{For simulating the two amplification stages of the readout board and the one of the oscilloscope, a transfer function of the circuit of the form $H(s) = 1/(\tau s + 1)$ was used in both cases, where $s$ represents the Laplace variable $s=j\omega$ (with $\omega$ being the radian frequency of the sinusoidal
signal) and $\tau$ represents the circuit time constant. For the amplification stages, the time constant was set to $\tau=RC$, with $R=50$ $\Omega$ and $C= 10$ pF, with the latter accounting for the detector and amplifier capacitance. For the oscilloscope stage, the time constant was set to $\tau=1/(2\pi f_c)$, with $f_c=1$ GHz. To simulate the total signal shaping effect, the $H(s)$ function for the two amplification stages and the oscilloscope are multiplied together with their respective $\tau$ values, while for the signal normalization, the value of the amplifier gain squared was used, i.e. $A^2 = (9.95)^2$ (see Appendix~\ref{app:amplifier_A}).}
The simulation of the electronics shows qualitatively the change of the sensor performance parameters and is not meant to provide a quantitative assessment of such effect. 
%the parametrization used in the WeightField2 and the readout parameters presented in Section~\ref{sec:DUT}. 
Figures~\ref{fig:Amp-FWHM-Risetime-Charge_simulations_LGAD} and~\ref{fig:Amp-FWHM-Risetime-Charge_simulations_Diode} show the values of the pulse amplitude, area, FWHM and rise time as functions of the bias voltage, simulated pre- and post-electronics for an LGAD and a diode, respectively, when impinged by a MIP and a 28-MeV proton. %as generated by the simulated sensors alone and read out by the simulated electronics.
The effect of the electronics has a significant impact on the values of FWHM and rise time.
%as the amplitude (defined as the peaks of the pulses), the pulse area (the integral of the pulse as a function of time), the full-width half-maximum (FWHM) and the rise time for the LGAD and the Diode. The latter parameter is defined as the difference between the times when the pulse surpasses 40$\%$ and $90\%$ of its peak value. Such threshold values were optimized to minimize the effect of noise in the rise time measurement.  %Gains for the LGAD is also calculated. 
The signals generated by the 28-MeV beam have amplitudes  10 times (diode) and about 3.7 times (LGAD) higher than those generated by a MIP. The values of FWHM and rise time of the LGAD and the diode are very similar in the beams of 28-MeV protons and beta-particles, when the effect of the electronics is included in the simulation. Prior to including the effect of the electronics in the LGAD signal, while the FWHM values decrease for increasing values of bias voltage, the FWHM ratios of 28-MeV protons to beta-particles increase to $1.13$ at 215 V. Conversely, prior to including the effect of the electronics, the rise-time is greater in a beta-particle beam than in a 28-MeV proton beam by a factor close to 2 at 215 V.   
Table~\ref{tab:params_simulations} also shows the gains of an LGAD, calculated with the two methods: as the ratio of amplitudes and of the  pulse areas in the LGAD and in the diode. With both methods the simulation predicts a gain suppression close to a factor 3, in a 28-MeV proton beam compared to a MIP beam. It is also noteworthy that the two methods provide estimates of gains that differ by a factor of about 3, when the effects of the electronics are not simulated, with the area method providing the largest gain values. The difference becomes far smaller when the effects of the electronics are included.

%%%%%%%%%%%%%%%%
\begin{table}[htb]
\centering
\begin{tabular}{|p{4.5cm}||p{4.4cm}|p{4.3cm}|}
 \hline
  Parameter& MIP                      &28-MeV Proton \\     
           &  LGAD (Diode)            &LGAD (Diode)\\
           &  Pre- / Post-Electronics  & Pre- / Post-Electronics\\
 \hline
 %\multicolumn{3}{|c|}{LGAD Simulation} \\
 \hline
 Amplitude [$\mu$A] / [mV]   & 9.4 (2.1) / 23 (2.3)  &32 (21)  / 84 (22) \\
 Area [$\mu$A $\cdot$ ns] / [mV $\cdot$ ns] &4.8 (0.42) / 24 (2.1) &18 (4.2) / 91 (21) \\
 FWHM [ns] &   0.53 (0.19) / 0.91 (0.79)  & 0.60 (0.19) / 0.94 (0.80) \\
 Rise time [ps] &230 (12) / 250 (170) & 120 (12) / 270 (170)\\
 \hline
 Gain by Amplitude Method & 4.4 / 9.9  & 1.5 / 3.8 \\
 Gain by Area Method    & 11 / 11  & 4.3 / 4.3\\ 
 \hline
\end{tabular}
\caption{Simulation predictions for amplitude, FWHM, and rise time for a diode and an LGAD in beams of MIPs and 28-MeV protons with the sensors biased at 215 V. The LGAD gains are calculated with two methods. The parameters are reported prior to and after the simulation of the effect of the readout electronics.}
\label{tab:params_simulations}
\end{table}
\FloatBarrier 
%%%
\begin{figure}[tb]
% \centering
      \setkeys{Gin}{width=\linewidth}
     \begin{subfigure}[t]{0.4\textwidth}
         \includegraphics[width=1.15\textwidth]{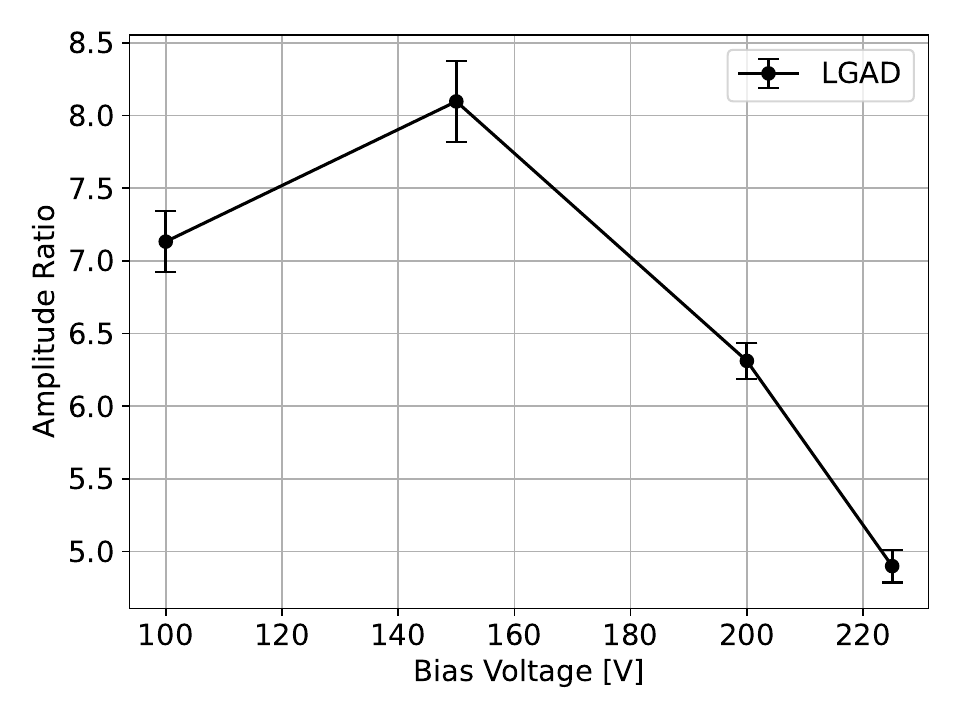}%{Picture1}
         \caption{Amplitude Ratios}
         \label{fig:AmplitudeRatio}
     \end{subfigure}
     \hfill
     \begin{subfigure}[t]{0.4\textwidth}
         %\centering
         \hspace{-1cm}
         \includegraphics[width=1.15\textwidth]{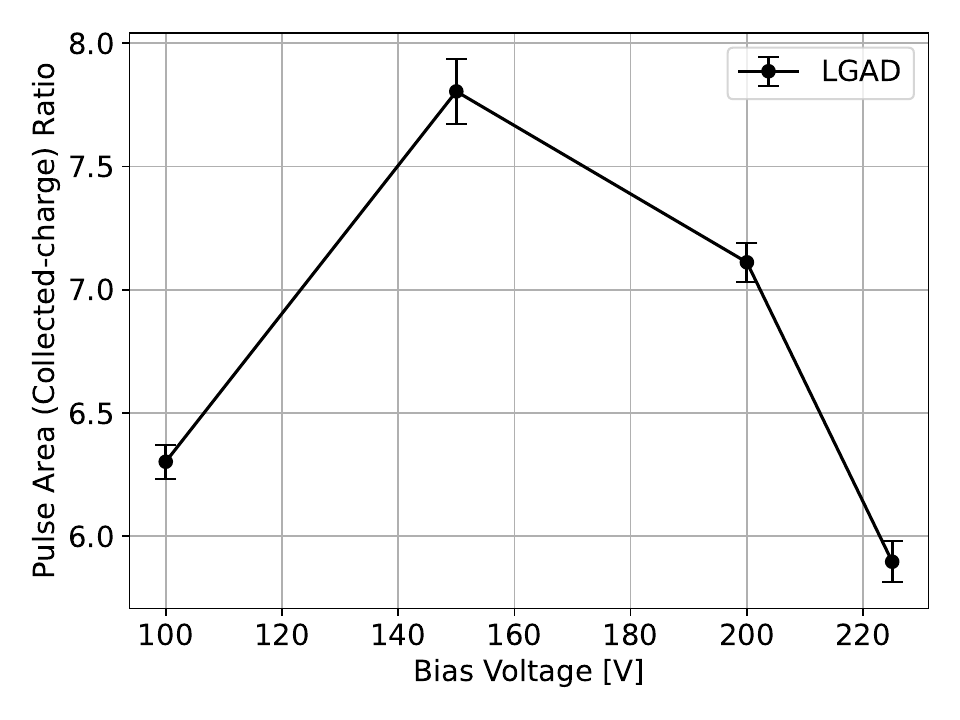}%{Picture3}
         \caption{Area Ratios}
         \label{fig:RiseTimeRatio}
     \end{subfigure}
     \hfill
     \begin{subfigure}[t]{0.4\textwidth}
         \centering
         \includegraphics[width=1.15\textwidth]{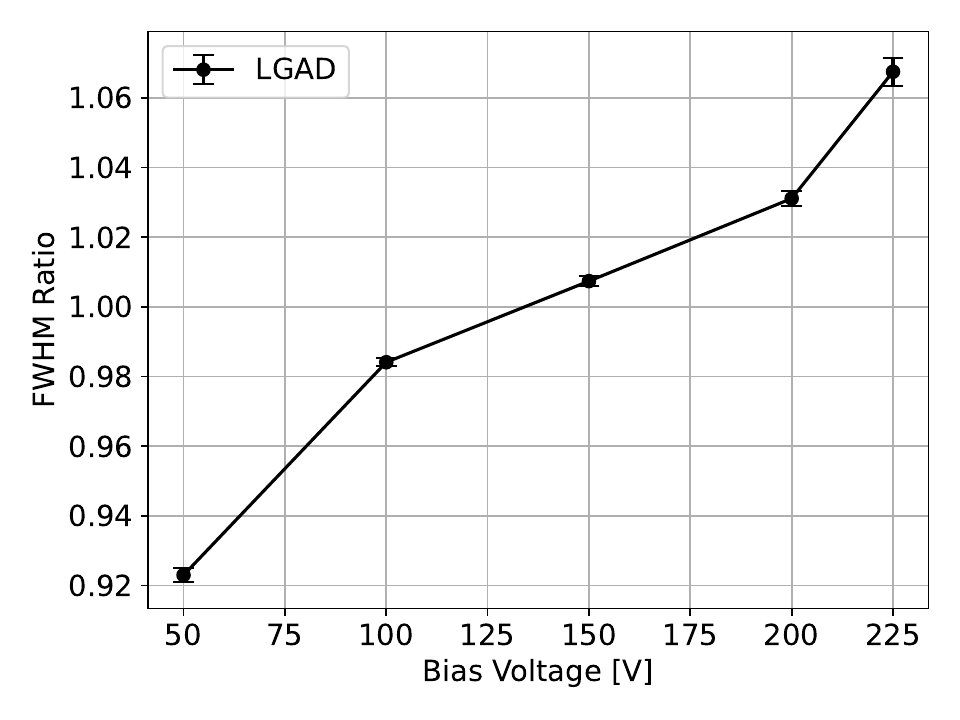}%{Picture2}
         \caption{FWHM Ratios}
         \label{fig:FWHMRatio}
     \end{subfigure}
     \hfill
     \begin{subfigure}[t]{0.4\textwidth}
         %\centering
        \hspace{-1cm}
         \includegraphics[width=1.15\textwidth]{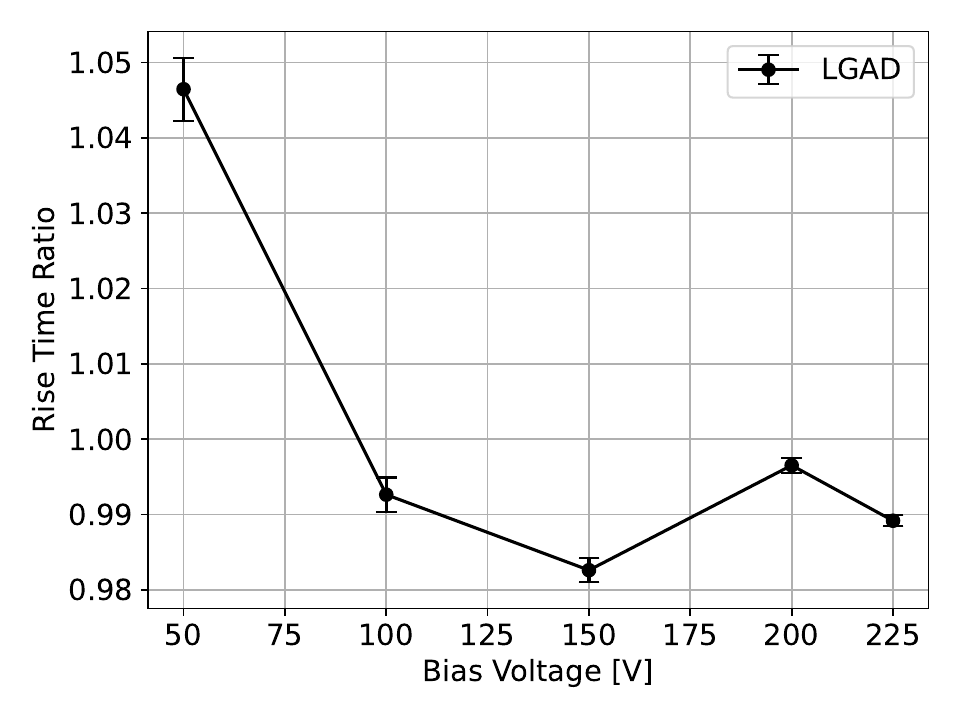}%{Picture3}
         \caption{Rise time Ratios}
         \label{fig:RiseTimeRatio}
     \end{subfigure}
     \caption{Ratios of amplitude (a), area (b), FWHM (c) and rise time (d)  in 28-MeV protons relative to beta-particles as functions of bias voltage, for the LGAD in data. The data points at 50 V bias voltage are not shown in (a) and (b), because the associated pulse amplitudes and areas for the beta-particle beams in data are truncated due to the triggering threshold in the oscilloscope (see Figures~\ref{fig:HistogramAmplitudeLGADs},~\ref{fig:HistogramAreaLGADs}).%The uncertainties are from the quadrature addition.
     }
    \label{fig:LGAD_ratios}
\end{figure}
%%%
%%%
\begin{figure}[t]
\centering
\includegraphics[width=.7\textwidth]{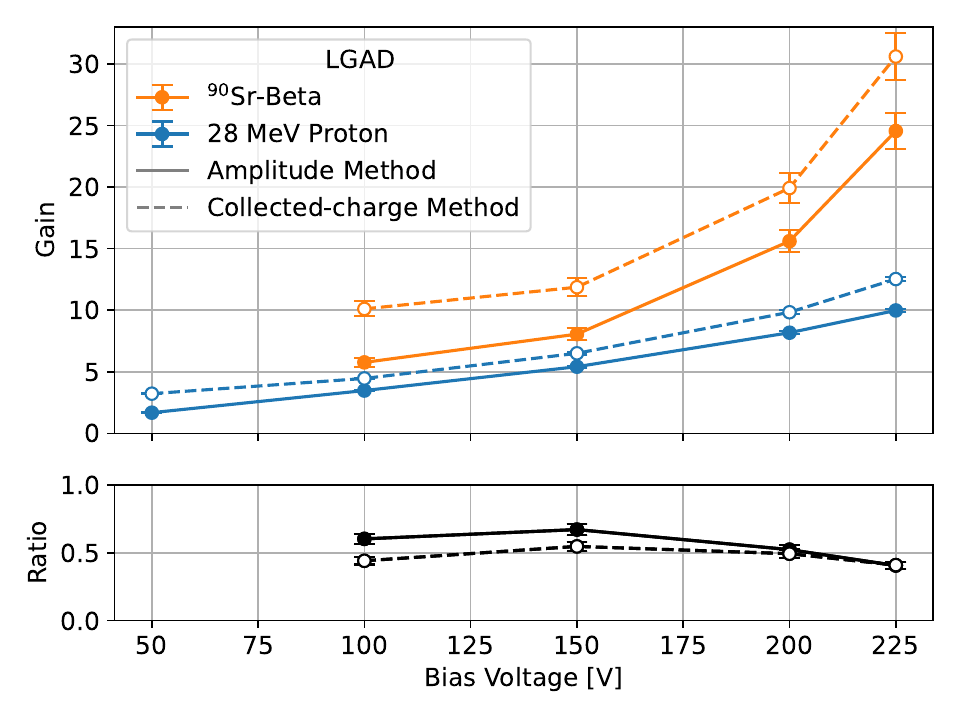}
\caption{Gains measured using the amplitude method and the collected-charge method for the LGAD in data and in simulation as functions of bias voltage in beta-particle and 28-MeV proton beams. At the bottom the ratios of the gain in the 28-MeV proton beam to the one in the beta-particle beam are shown for each value of the bias voltage for the amplitude method and the collected-charge method. The data points at 50 V bias voltage are not shown for the beta-particle beams for the gains and in the ratios, because the associated pulse amplitudes and areas in data are truncated due to the triggering threshold in the oscilloscope (see Figures~\ref{fig:HistogramAmplitudeLGADs},~\ref{fig:HistogramAreaLGADs}).
}
\label{fig:GainSummary_BothMethods}
\end{figure}
%%%

%%%%%%%%%%%%%%%%%%%%%%%%%%%%%%%%%%%%%%%%%%%%%%%%%
\section{Comparison of results}  
\label{sec:comparison}
%%%%%%%%%%%%%%%%%%%%%%%%%%%%%%%%%%%%%%%%%%%%%%%%%

The response of the LGAD in the 28-MeV proton beam is compared to that in the $^{90}{\rm Sr}$-beta beam.  Ratios of parameters measured in data with the 28-MeV proton beam and the $^{90}{\rm Sr}$-beta beam are provided as well as those predicted by the simulation.

Figure~\ref{fig:LGAD_ratios} shows the experimental results for the ratios of amplitude, area, FWHM and rise time. %The ratios of amplitudes vary as functions of bias voltage with a similar dependence in both LGADs.
The ratios of pulse amplitudes and areas are very similar.
When the sensor is under a high bias voltage, i.e.\ at or greater than 100 V, the amplitudes and areas in a 28-MeV proton beam are in the range of 5-8 times larger than in a beta-particle beam, in agreement with the simulation that predicts the ratios to be approximately 3-10, regardless of whether the effects of the electronics are included or not. For bias voltages $\ge 100$ V, the values of FWHM and rise time are stable as functions of bias voltage and very close to 1, in agreement with the simulation, with and without the effects of the electronics. 
%The duration of the pulses (FWHM) is slightly longer in the  28-MeV proton beam than in the beta-particle beam by a factor up to approximately 1.07.  The factor is about 1.1 in the simulation after the effects of the electronics are included. 
Similarly, the speed of the signal (rise time) is consistent in the two beam types in data, in agreement with the simulation when the effects of the electronics are included, whereas, without the effect of the electronics, the simulation predicts it to be faster in the 28-MeV proton beam by a factor of approximately 2 at 215 V and approximately 10 or greater for lower values of bias voltage. 

Strong linear correlation between the pulse amplitudes and areas is shown in Figure~\ref{fig:AreaVsAmplitudes}. This justifies the common practice to use either variable for the measurement of the gain of an LGAD. The LGAD gains measured using the two methods are compared in Figure~\ref{fig:GainSummary_BothMethods} for the $^{90}{\rm Sr}$-beta beam and the 28-MeV proton beam.
Significant systematic differences are observed between the two methods for measuring gain, with the charge-collection method giving larger gain values than the amplitude method.
The difference between the gains measured with the two methods is approximately 1.5 (28-MeV protons) and 5.0  (beta particles) for bias voltages greater than 100 V. While simulation predicts different values of gains with respect to those measured in data, it also predicts a systematic difference between the two methods for measuring the gain, but only when the effects of the electronics are not accounted for. When the effects of electronics are included in the simulation, the difference between the gains measured with the two methods is largely reduced.  

 Despite the systematic differences between the two methods in the measured values of gains, they both agree on the general observation of gain suppression in a 28-MeV proton beam compared to a beta-particle beam.  
 The data show the gain suppression to be dependent on the bias voltage and to be about 2.5 with both gain-calculation methods at a bias voltage of 225 V. The simulation also predicts gain suppression of about 2.6-3.0 with both gain-calculation methods, with and without the effects of the electronics.
 %For LGAD-2 the value gain suppression depends on the gain method, and is about 2 with the amplitude method and about 1.3 with the collected-charge method at a bias voltage of 225 V.  
Such a gain suppression can occur in silicon sensors with different particle beams and depends on the amount of deposited charge~\cite{GIACOMINI2024169605}. For charge particles impinging a silicon detector, the charges created by ionization are distributed along a track. A self-screening effect of the charges occurs when large clouds of charges are created in a small volume: the inner carriers in the charge cloud are shielded from the external electric field by the outermost charges, which furthermore lowers the local magnitude of the electric field. As a result, the charge carriers experience a limited multiplication. 

%%%%%%%%%%%%%%%%%%%%%%%%%%%%%%%%%%%%%%%%%%%%%%%%%
\section{Conclusions}  
\label{sec:conclusions}

The performance of an LGAD was studied in a beam of 28-MeV protons and compared to that in a MIP beam from $^{90}{\rm Sr}$ beta  particles and in simulation. Pulse amplitude, width, rise time and gain, were studied as functions of the bias voltage. Qualitatively good agreement is found between data and simulation. %, but a significant difference in response is observed between the two LGADs, due to differences in fabrication. 
The pulse amplitude and area are significantly larger in the 28-MeV proton beam than in the beta-particle beam, by a factor of approximately 2-8. The widths and rise time show instead smaller differences between the two particle beams. To measure the gain, two methods were used, one based on pulse amplitudes and the other on the collected charge. The two methods result in a systematic difference in the measured gains, with the one based on the collected charge yielding larger gain values in data. Despite such differences, gain suppression is observed in the LGAD with a 28-MeV proton beam with respect to a MIP beam. As for the other measured parameters, the gain suppression also depends on the bias voltage and is about 2.5 at high bias voltage. This set of results will be valuable for the use of LGAD-based sensors in non-MIP beams for applications beyond high energy or nuclear physics, such as low-energy rare processes, medical, biological, and chemical, in which beams of non-MIPs are used.

%%%%%%%%%%%%%%%%%%%%%%%%%%%%%%%%%%%%%%%%%%%%%%%%%

%%%%%%%%%%%%%%%%%%%%%%%%%%%%%%%%%%%%%%%%%%%%%%%%%%%%%
\appendix
\section{Measurement of the number of electron-hole pairs generated by a MIP in silicon}
\label{app:n_ehPairs_meas}

\begin{figure}[bth]
\centering

\includegraphics[width=.45\textwidth]{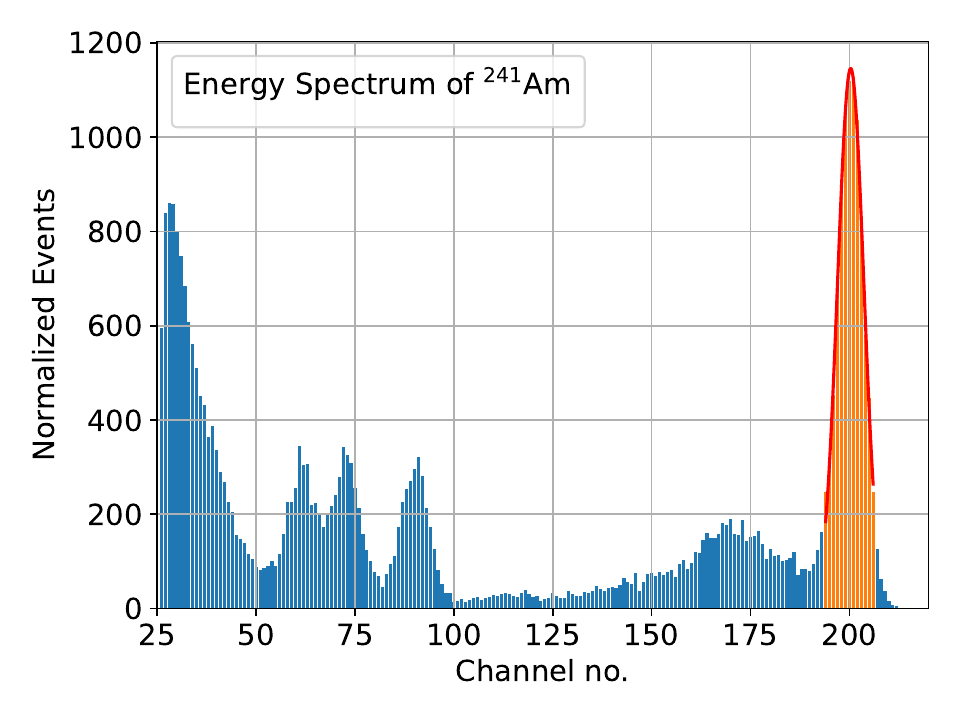}
\includegraphics[width=.45\textwidth]{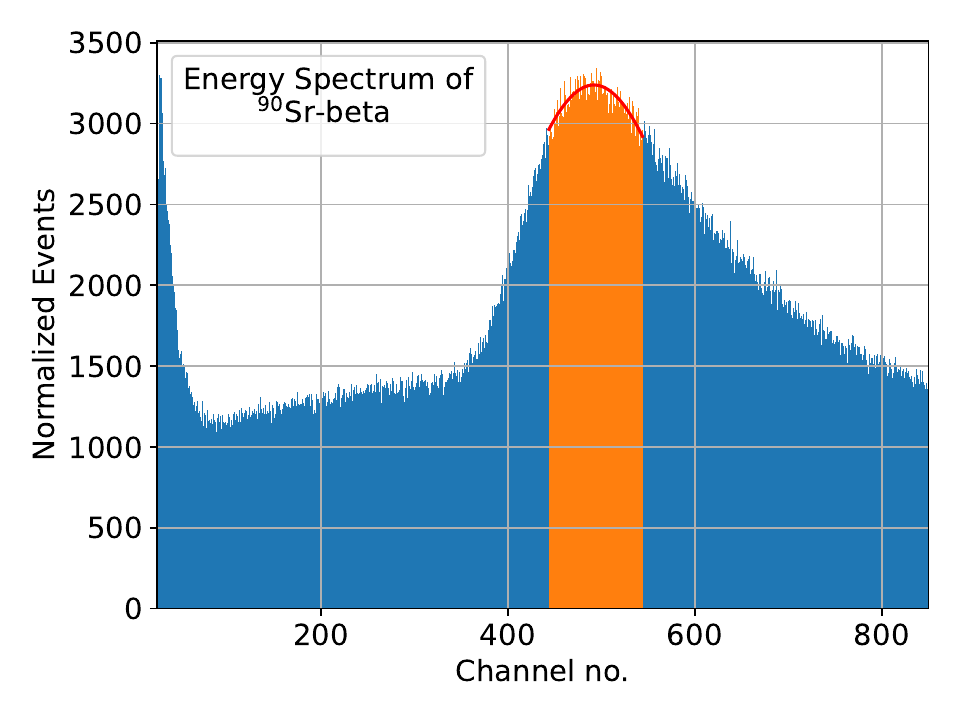}
\caption{Energy spectra produced in a diode of 500 $\mu$m  active thickness by X-rays from an $^{241}{\rm Am}$ radioactive source (left) and a beta-beam from a $^{90}$Sr radioactive source (right). The result of the Gaussian fits around the 59.5 keV (left) and 146 keV (right) peaks are also shown, and the fit areas are colored in orange.
}
\label{fig:AmSpectrum}
\end{figure}

The number of electron-hole pairs generated by a MIP in silicon was measured using a diode of 500~$\mu$m  active thickness exposed to a beam of beta-particles from a $^{90}{\rm Sr}$  radioactive source. The sensor was read out by a charge-sensitive amplifier and the PX5 (by Amptek Inc.) readout system. The calibration of the number of readout channel counts as a function of the energy was carried out with X-rays from a $^{241}{\rm Am}$ radioactive source. Figure~\ref{fig:AmSpectrum} (left) shows the X-ray spectrum and the peak corresponding to the 59.5 keV spectrum line. The peak of the beta spectrum was found in the readout channel (channel no.\ 494) corresponding to 146 keV, see Figure~\ref{fig:AmSpectrum} (right). Dividing this value by 3.6 eV, i.e.\ the energy necessary to generate an electron-hole pair in silicon, the number of electron-hole pairs generated in the 500~$\mu$m thickness of silicon is found to be $40.6 \times 10^3$, which corresponds to $81.20 \pm 0.16$ electron-hole pairs in one micron of silicon. 

The number of electron-hole pairs in a silicon sensor is dependent on the thickness of the active volume~\cite{RevModPhys.60.663}. The predicted value of such a parameter in the model presented in Ref.~\cite{RevModPhys.60.663} (Table V) for a silicon of 500~$\mu$m thickness was found to be compatible with the value measured experimentally and presented above. Thus, the same model was used to extrapolate $81.20 \pm 0.16$ for a 500 $\mu$m thickness to $62.9 \pm 3.8$ for a 35~$\mu$m thickness, where the uncertainty on the extrapolation model is conservatively increased by a factor of 5 with respect to the quoted $1.2\%$, because the value of the $\beta\gamma$ parameter from the beta-particle beam from the $^{90}{\rm Sr}$ radioactive source is smaller than the one that was assumed in the model. 

%%%
%%%

%%%%%%%%%%%%%%%%%%%%%%%%%%%%%%%%%%%%%%%%%%%%
\section{Measurement of the Amplifier Gain}
\label{app:amplifier_A}

\begin{figure}[tb]
\centering
\includegraphics[width=.6\textwidth]{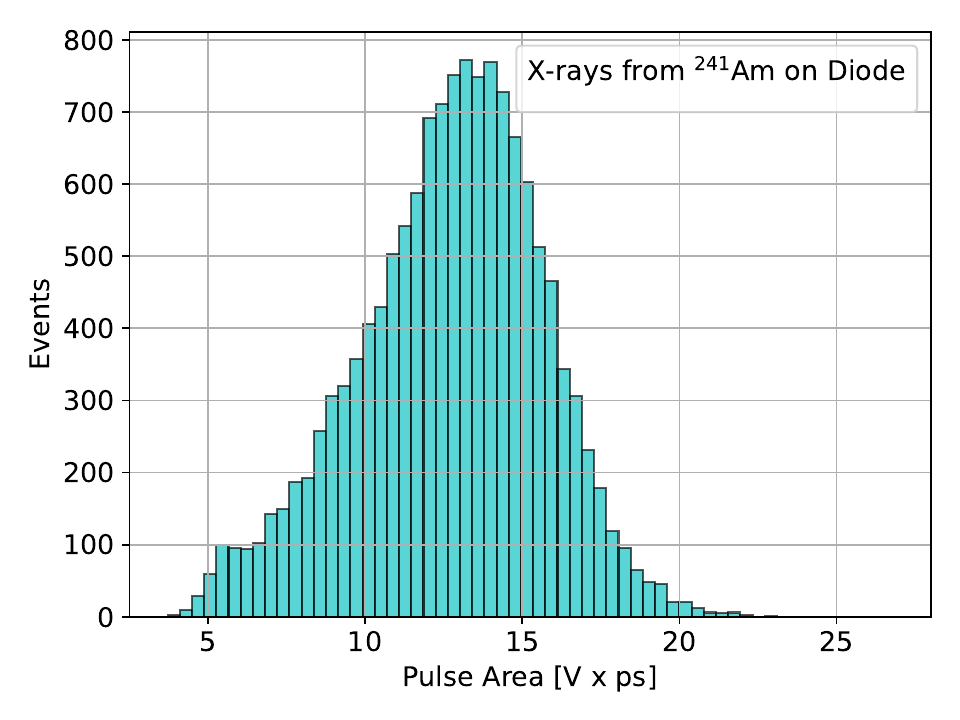}
\caption{Distribution of the integral (area) of pulses as a function of time for $^{241}$Am at 100~V, after noise rejection. 
}
\label{fig:AmAreaHistogram}
\end{figure}
%%%

The amplifier gain in the FNAL board was measured using a diode fabricated by HPK with the same design as those used in the main body of this paper, mounted on the FNAL board. Data were collected with a beam of X-rays from an $^{241}$Am radioactive source.  

The pulse measured in an oscilloscope can be expressed as a function of the signal current as follows:
\begin{equation}
    V_{\rm scope} = i_{\rm signal} \cdot (50~{\rm Ohm})  \cdot A^2,
\end{equation}
where $A$ is the amplifier gain, which is squared in the above equation due to the two-stage amplification present in the FNAL board. 
By integrating each term in the above equation over time, we obtain the following:
\begin{equation}
    \int dt V_{\rm scope} = Q \cdot (50~{\rm Ohm})  \cdot A^2,
\end{equation}
where $Q$ is the measured charged.
From the distribution of the integral of the pulses as a function of time, as shown in Figure~\ref{fig:AmAreaHistogram}, the most probably value is measured as $\int dt V_{\rm scope} = (12.92 \pm 0.06) \times 10^{-12}$ V ps from a Gaussian fit to data.

The number of electron-hole pairs generated in the sensor is estimated to be 16.6 k by dividing by 3.6 eV the 59.5 keV energy of the X-ray emission from the $^{241}$Am decay. The charge $Q$ is then estimated by multiplying 16.6 k by the electron charge. Thus, the value of the amplification gain is extracted to be $A=9.95\pm 0.02$, in agreement with the value reported in the device data-sheet, i.e.\ the  Mini-Circuits GALI-66+ integrated circuit~\cite{MiniCircuitsGALI66}.

As an additional cross-check, the proportionality constant between the area and the amplitude of a pulse generated by X-rays from the $^{241}$Am source is compared to the one extracted in Sec.~\ref{sec:gain_ampl_beta} for a MIP from a ${}^{90}$Sr source, by extrapolating a linear fit to data in a two-dimensional area-amplitude distribution. For the latter, the area-to-amplitude proportionality constant is $0.66\pm 0.01$ ns, while for the former the most probable values of the amplitude and area distributions are 
$19.0 \pm 0.1$~mV  and $12.92 \times 10^{-12}$ V ps (as reported above), respectively, leading to a proportionality constant of 0.68 $\pm$ 0.01 ns. The area-to-amplitude proportionality constant can be considered as an effective FWHM. 
Their values measured with X-rays and a MIP are in agreement and are also found to be compatible with the FWHM values shown in Figure~\ref{fig:HistogramFWHMLGADs} for an LGAD in a beta-particle beam from a ${}^{90}$Sr source as well as for a diode in a 28-MeV proton beam, as shown in Figure~\ref{fig:Diode_FHWM}.

%%%%%%%%%%%%%%%%%%%%%%%%%%%%%%%%%%%%%%%%%%%%%
\clearpage

%%%%%%%%%%%%%%%%%%%%%%%%%%%%%%%%%%%%%%%%%%%%%
\acknowledgments

The authors wish to thank their colleagues at Brookhaven National Laboratory: Don Pinelli, Antonio Verderosa, Joe Pinz and Tim Kersten for sensor mounting. This material is based upon work supported by the U.S.\ Department of Energy under grants DE-SC0012704, DE-SC443363, DE-SC426496 and DE-SC0020255. %This research used resources of the Center for Functional Nanomaterials, which is a U.S. DOE Office of Science Facility, at Brookhaven National Laboratory under Contract No. DE-SC0012704.

%\paragraph{Note added.} This is also a good position for notes added after the paper has been written.

% Bibliography

%% [A] Recommended: using JHEP.bst file
\bibliographystyle{JHEP}
\bibliography{biblio.bib}

\providecommand{\href}[2]{#2}\begingroup\raggedright\begin{thebibliography}{10}

\bibitem{CMS:2667167}
{CMS Collaboration}, \emph{{A MIP Timing Detector for the CMS Phase-2 Upgrade}},  Tech. Rep. \href{https://cds.cern.ch/record/2667167}{CERN-LHCC-2019-003. CMS-TDR-020}, CERN, Geneva (Mar, 2019).

\bibitem{Collaboration:2623663}
{ATLAS Collaboration}, \emph{{Technical Proposal: A High-Granularity Timing Detector for the ATLAS Phase-II Upgrade}},  Tech. Rep. \href{http://cds.cern.ch/record/2623663}{CERN-LHCC-2018-023. LHCC-P-012}, CERN, Geneva (Jun, 2018).

\bibitem{CERN-ACC-2015-0140}
I.~Bejar~Alonso and L.~Rossi, \emph{{HiLumi LHC Technical Design Report: Deliverable: D1.10}},  Tech. Rep. \href{https://cds.cern.ch/record/2069130}{CERN-ACC-2015-0140} (Nov, 2015).

\bibitem{CERN-ATS-2012-236}
L.~Rossi and O.~Bruning, \emph{{High Luminosity Large Hadron Collider: A description for the European Strategy Preparatory Group}},  Tech. Rep. \href{https://cds.cern.ch/record/1471000}{CERN-ATS-2012-236} (Aug, 2012).

\bibitem{hartmut}
H.F.W.~Sadrozinski, S.~Ely, V.~Fadeyev, Z.~Galloway, J.~Ngo, C.~Parker et~al., \emph{Ultra-fast silicon detectors}, \href{https://doi.org/10.1016/j.nima.2019.04.073}{\emph{Nucl. Inst. Meth. A} {\bfseries 730} (2013) 226}.

\bibitem{GIACOMINI201952}
G.~Giacomini, W.~Chen, F.~Lanni and A.~Tricoli, \emph{{Development of a technology for the fabrication of Low-Gain Avalanche Diodes at BNL}}, \href{https://doi.org/10.1016/j.nima.2013.06.033}{\emph{Nucl. Inst. Meth. A} {\bfseries 934} (2019) 52}.

\bibitem{micronlgad}
N.~Moffat, R.~Bates, M.~Bullough, L.~Flores, D.~Maneuski, L.~Simon et~al., \emph{Low gain avalanche detectors ({LGAD}) for particle physics and synchrotron applications}, \href{https://doi.org/10.1088/1748-0221/13/03/C03014}{\emph{JINST} {\bfseries 13} (2018) C~03014}.

\bibitem{Lange:2017pxs}
J.~Lange et~al., \emph{{Gain and time resolution of 45 $\mu$m thin Low Gain Avalanche Detectors before and after irradiation up to a fluence of $10^{15}$ n$_{eq}$/cm$^2$}}, \href{https://doi.org/10.1088/1748-0221/12/05/P05003}{\emph{JINST} {\bfseries 12} (2017) P05003} [\href{https://arxiv.org/abs/1703.09004}{{\ttfamily 1703.09004}}].

\bibitem{Zhao:2018qkg}
Y.~Zhao et~al., \emph{{Comparison of 35 and 50 $\mu$m thin HPK UFSD after neutron irradiation up to 6*10$^{15}$ neq/cm$^2$}}, \href{https://doi.org/10.1016/j.nima.2018.08.040}{\emph{Nucl. Inst. Meth. A} {\bfseries 924} (2019) 387} [\href{https://arxiv.org/abs/1803.02690}{{\ttfamily 1803.02690}}].

\bibitem{Li:2021iid}
M.~Li et~al., \emph{{The Timing Resolution of IHEP-NDL LGAD Sensors With Different Active Layer Thicknesses}}, \href{https://doi.org/10.1109/TNS.2021.3097746}{\emph{IEEE Trans. Nucl. Sci.} {\bfseries 68} (2021) 2309}.

\bibitem{YANG2020164379}
X.~Yang, S.~Alderweireldt, N.~Atanov, M.~Ayoub, J.B.G.~{da Costa}, L.C.~Garcia et~al., \emph{{Layout and performance of HPK prototype LGAD sensors for the High-Granularity Timing Detector}}, \href{https://doi.org/https://doi.org/10.1016/j.nima.2020.164379}{\emph{Nucl. Inst. Meth. A} {\bfseries 980} (2020) 164379}.

\bibitem{DUTTA2025170224}
I.~Dutta, C.~Madrid, R.~Heller, S.~Nanda, D.~Shekar, C.S.~Mart\'in et~al., \emph{{Results for pixel and strip centimeter-scale AC-LGAD sensors with a 120 GeV proton beam}}, \href{https://doi.org/https://doi.org/10.1016/j.nima.2025.170224}{\emph{Nucl. Inst. Meth. A} {\bfseries 1072} (2025) 170224}.

\bibitem{ACLGADprocess}
G.~Giacomini, W.~Chen, G.~D'Amen and A.~Tricoli, \emph{{Fabrication and Performance of AC-coupled LGADs}}, \href{https://doi.org/10.1088/1748-0221/14/09/P09004}{\emph{JINST} {\bfseries 14} (2019) P09004}.

\bibitem{RSD_NIM}
M.~Mandurrino, R.~Arcidiacono, M.~Boscardin, N.~Cartiglia, G.F.~{Dalla Betta}, M.~Ferrero et~al., \emph{{Analysis and numerical design of Resistive AC-Coupled Silicon Detectors (RSD) for 4D particle tracking}}, \href{https://doi.org/10.1016/j.nima.2020.163479}{\emph{Nucl. Inst. Meth. A} {\bfseries 959} (2020) 163479}.

\bibitem{AbdulKhalek:2021gbh}
R.~Abdul~Khalek et~al., \emph{{Science Requirements and Detector Concepts for the Electron-Ion Collider}: {EIC Yellow Report}}, \href{https://doi.org/10.1016/j.nuclphysa.2022.122447}{\emph{Nucl. Phys. A} {\bfseries 1026} (2022) 122447} [\href{https://arxiv.org/abs/2103.05419}{{\ttfamily 2103.05419}}].

\bibitem{10.3389/fphy.2020.578444}
M.~Missiaggia, E.~Pierobon, M.~Castelluzzo, A.~Perinelli, F.~Cordoni, M.~Centis~Vignali et~al., \emph{{A Novel Hybrid Microdosimeter for Radiation Field Characterization Based on the Tissue Equivalent Proportional Counter Detector and Low Gain Avalanche Detectors Tracker: A Feasibility Study}}, \href{https://doi.org/10.3389/fphy.2020.578444}{\emph{Frontiers in Physics} {\bfseries 8} (2021) }.

\bibitem{instruments5040040}
S.M.~Mazza, \emph{{An LGAD-Based Full Active Target for the PIONEER Experiment}}, \href{https://doi.org/10.3390/instruments5040040}{\emph{Instruments} {\bfseries 5} (2021) }.

\bibitem{Tandem}
{Brookhaven National Laboratory}, \emph{{Tandem Van de Graaff}}, {\emph{https://www.bnl.gov/tandem/} (2024) }.

\bibitem{HELLER2021165828}
R.~Heller, A.~Abreu, A.~Apresyan, R.~Arcidiacono, N.~Cartiglia, K.~DiPetrillo et~al., \emph{{Combined analysis of HPK 3.1 LGADs using a proton beam, beta source, and probe station towards establishing high volume quality control}}, \href{https://doi.org/https://doi.org/10.1016/j.nima.2021.165828}{\emph{Nucl. Inst. Meth. A} {\bfseries 1018} (2021) 165828}.

\bibitem{MiniCircuitsGALI66}
{Mini-Circuits}, \emph{{GALI-S66+}}, {\emph{https://www.minicircuits.com/pdfs/GALI-S66+.pdf} }.

\bibitem{NISt}
{National Institute of Standards and Technology}, \emph{{The PSTAR program calculates stopping power and range tables for protons in various materials}}, {\emph{https://physics.nist.gov/PhysRefData/Star/Text/PSTAR.html} (2024) }.

\bibitem{GIACOMINI2024169605}
G.~Giacomini, W.~Chen, G.~D’Amen, E.~Rossi and A.~Tricoli, \emph{Spectroscopic performance of low-gain avalanche diodes for different types of radiation}, \href{https://doi.org/https://doi.org/10.1016/j.nima.2024.169605}{\emph{Nuclear Instruments and Methods in Physics Research Section A: Accelerators, Spectrometers, Detectors and Associated Equipment} {\bfseries 1066} (2024) 169605}.

\bibitem{RevModPhys.60.663}
H.~Bichsel, \emph{Straggling in thin silicon detectors}, \href{https://doi.org/10.1103/RevModPhys.60.663}{\emph{Rev. Mod. Phys.} {\bfseries 60} (1988) 663}.

\end{thebibliography}\endgroup

%% or
%% [B] Manual formatting (see below)
%% (i) We suggest to always provide author, title and journal data or doi:
%% in short all the informations that clearly identify a document.
%% (ii) please avoid comments such as "For a review'', "For some examples",
%% "and references therein" or move them in the text. In general, please leave only references in the bibliography and move all
%% accessory text in footnotes.
%% (iii) Also, please have only one work for each \bibitem.

%AT:
%\begin{thebibliography}{99}

%\bibitem{a}
%Author,
%\emph{Title},
%\emph{J. Abbrev.} {\bf vol} (year) pg.

%\bibitem{b}
%Author,
%\emph{Title},
%arxiv:1234.5678.

%\bibitem{c}
%Author,
%\emph{Title},
%Publisher (year).

%\end{thebibliography}

\end{document}